\documentclass[aps,showpacs,preprintnumbers,superscriptaddress,merge,notitlepage]{revtex4-1}
\usepackage{hyperref,amsmath,subfigure,color}
\usepackage{graphicx}

\begin{document}
\title{Fluctuation-dissipation dynamics of cosmological scalar fields} 
\author{Sam Bartrum}
\email{s.bartrum@sms.ed.ac.uk}

\author{Arjun Berera}
\email{ab@ph.ed.ac.uk}
\affiliation{SUPA, School of Physics and Astronomy, University of Edinburgh, Edinburgh, EH9 3JZ, United Kingdom}

\author{Jo\~ao G. Rosa\footnote{Also at Departamento de F\'{\i}sica e Astronomia, Faculdade de Ci\^encias da Universidade do Porto, Rua do Campo Alegre 687, 4169-007 Porto, Portugal.}}
\email{joao.rosa@ua.pt}
\affiliation{Departamento de F\'{\i}sica da Universidade de Aveiro and CIDMA, Campus de Santiago, 3810-183 Aveiro, Portugal}

%\preprint{[TO DO]}

%%%%%%%%%%%%%%%%%%%%%%%%%%%%%%%%%%%%%%%%%%%%%%%%%%%%%%%%%%%%%%%%%%%%%%%%%%%%%%%%%%%%%%%%%%%%%%%%%%%%%%%%%%%%%%%%%%%%%%%%%%%%%%%%%%%%%%%%
%%%%%%%%%%%%%%%%%%%%%%%%%%%%%%%%%%%%%%%%%%%%%%%%%%%%%%%%%%%%%%%%%%%%%%%%%%%%%%%%%%%%%%%%%%%%%%%%%%%%%%%%%%%%%%%%%%%%%%%%%%%%%%%%%%%%%%%%

\begin{abstract}
We show that dissipative effects have a significant impact on the evolution of cosmological scalar fields, leading to friction, entropy production and field fluctuations. We explicitly compute the dissipation coefficient for different scalar fields within the Standard Model and some of its most widely considered extensions, in different parametric regimes. We describe the generic consequences of fluctuation-dissipation dynamics in the post-inflationary universe, focusing in particular on friction and particle production, and analyze in detail two important effects. Firstly, we show that dissipative friction delays the process of spontaneous symmetry breaking and may even damp the the motion of a Higgs field sufficiently to induce a late period of warm inflation. Along with dissipative entropy production, this may parametrically dilute the abundance of dangerous thermal relics. Secondly, we show that dissipation can generate the observed baryon asymmetry without symmetry restoration, and we develop in detail a model of dissipative leptogenesis. We further show that this generically leads to characteristic baryon isocurvature perturbations that can be tested with CMB observations. This work provides a fundamental framework to go beyond the leading thermal equilibrium semi-classical approximation in addressing fundamental problems in modern cosmology.

\end{abstract}

\maketitle

%\tableofcontents

%%%%%%%%%%%%%%%%%%%%%%%%%%%%%%%%%%%%%%%%%%%%%%%%%%%%%%%%%%%%%%%%%%%%%%%%%%%%%%%%%%%%%%%%%%%%%%%%%%%%%%%%%%%%%%%%%%%%%%%%%%%%%%%%%%%%%%%%
%%%%%%%%%%%%%%%%%%%%%%%%%%%%%%%%%%%%%%%%%%%%%%%%%%%%%%%%%%%%%%%%%%%%%%%%%%%%%%%%%%%%%%%%%%%%%%%%%%%%%%%%%%%%%%%%%%%%%%%%%%%%%%%%%%%%%%%%

\section{Introduction}

Scalar fields play a major role in modern cosmological theories. Depending on the balance between the kinetic, potential and gradient energy stored in these fields, they can mimic fluids with distinct equations of state and so have been proposed as leading candidates to describe the early phase of inflationary expansion, as well as dark matter and dark energy. 

Scalar fields are also a key ingredient in modern particle physics theories and the recent discovery of what is now widely accepted to be the electroweak Higgs boson at the LHC puts the existence of fundamental scalar degrees of freedom on firm experimental ground. Indeed, these are ubiquitous in extensions of the Standard Model (SM) of particle physics, such as grand unified theories, supersymmetric theories or extra-dimensional scenarios, namely within the context of string/M-theory compactifications. The study of the cosmological dynamics of scalar fields, both at the classical and quantum levels, is thus of crucial importance to understand the early history of our universe.

One of the most prominent roles of scalar fields is the phenomenon of spontaneous symmetry breaking in fundamental gauge theories, where vector bosons and fermions acquire mass through the Bose-condensation of a scalar field. This process of spontaneous symmetry breaking sees an initial symmetric state go to a state of broken symmetry, all due to the change of a single parameter, the vacuum expectation value (vev) of a scalar field $\langle \phi \rangle$. The electroweak Higgs mechanism is the best known example of this simple idea, which is also expected to apply to the spontaneous breaking of higher-rank gauge symmetry groups that extend the SM at high energy scales.

The significance of spontaneous symmetry breaking for cosmology was pointed out several decades ago by Kirzhnits and Linde \cite{Kirzhnits:1976ts}. They observed that this behavior would be a feature of quantum field theories at finite temperature, whereby at very high temperatures the vev of the scalar field would be a single value that restores symmetry and then, at some specific critical temperature, this vev would change and lead to a phase of broken symmetry. The description of this process fits well within the Landau theory of phase transitions. These two simple ideas of spontaneous symmetry breaking and its realization in finite temperature quantum field theory as a phase transition have been the foundation for cosmological phase transitions \cite{Linde:1978px,Linde:1981zj}. Such behavior has since been applied to numerous areas in cosmology including inflation, defects, baryogenesis \cite{Dine:1992wr}, and cosmic magnetic fields \cite{Vachaspati:1991nm}.

The study of cosmological phase transitions has so far been centered primarily on their equilibrium properties. In particular, most of the interest has gone into studying the particle physics features in the symmetric and broken phases. The dynamics that induces the change from one phase to the other is, however, also a necessary component of this entire process. This change will involve the motion of the order parameter $\langle \phi \rangle$ from the symmetry-restored to broken phase.  Since this scalar order parameter is the expectation value of a quantum field, which in general interacts with other fields that comprise the radiation bath, its evolution between the different phases will generically involve energy exchange. Due to the tendency for the equipartion of energy in dynamical systems, this appears primarily as energy exchange between the single dynamical scalar degree of freedom and the many degrees of freedom comprising the heat bath. This thus results in dissipation of the scalar field's energy into the ambient radiation fluid.

The order parameter experiences another effect when immersed in the radiation bath. All the random interactions of the bath constituents with the single order parameter will slightly push the scalar field around in all different directions, thus inducing fluctuations. These two processes of dissipation and fluctuations of the order parameter are intrinsically related to each other by the underlying dynamical quantum mechanical equations \cite{Callen:1951vq}. This is the basis of fluctuation-dissipation theorems and it is applicable to the dynamics of cosmological phase transitions just as it is to any phase transitions or out-of-equlibrium situation in condensed matter systems \cite{Negele:1988vy, citeulike:8167873, Caldeira:1982uj,Goldenfeld:1992qy}.

As such, wherever a cosmological phase transition is present, fluctuation-dissipation dynamics will be present hand- in-hand during the out-of-equilibrium transition period between the two equilibrium phases. This phase transition dynamics will add three new features to the equilibrium description. First, the background evolution of the scalar order parameter will affect the expansion behaviour of the Universe. Second, there will be particle production. Third, there will be fluctuations created in the Universe in the wake of this transition. The first feature has been examined in great detail in the cosmology literature. For the last two features, there are many quantum field theory calculations of fluctuation and/or dissipation dynamics \cite{Berera:1998gx,Boyanovsky:1996rw,Boyanovsky:1997xt,calzetta2008nonequilibrium,Gleiser:1993ea, PhysRevD.67.045006, PhysRevD.60.063510, PhysRevD.61.025012, Ringwald:1987ui,Hosoya:1983ke, Morikawa:1986rp, Gleiser:1993ea} but very little has been explicitly applied to particle physics models during cosmological phase transitions. One exception is in the case of inflation, where warm inflation captures all three of these features \cite{Bartrum:2013fia,Bartrum:2013oka,Bartrum:2012tg,Bastero-Gil:2014oga,Berera:1995ie,Berera:1996nv,Berera:1999ws,Hall:2003zp,Moss:2008yb,BasteroGil:2009ec,Berera:2002sp,Berera:2008ar}. However, cosmological phase transitions can and generically do occur with no inflation  and, in these cases also, all three of these features will be present. They are an intrinsic part of the evolution history of the early Universe and the dynamics emerging from whatever is the underlying particle physics model.

Fluctuation-dissipation effects will, more generally, be present in the dynamics of any cosmological scalar field, regardless of the occurrence of phase transitions. For example, several completions of the SM predict the existence of very light scalars, such as extra-dimensional moduli or axion-like fields. These fields will be underdamped during the early inflationary phase and driven to potentially very large values by random quantum fluctuations. After inflation, once the expansion rate has decreased sufficiently, they will be able to dynamically relax to their minimum energy configuration. This may in several cases lead to large-amplitude oscillations that overclose the universe or spoil the successful predictions of Big Bang Nucleosynthesis (BBN) for light element abundances, which poses a considerable challenge for cosmological models in beyond the SM scenarios. Interactions with other degrees of freedom in the ambient heat bath may, however, induce energy dissipation and fluctuations in these scalar fields, modifying their dynamical evolution and potentially their role in the subsequent cosmic history.

Another case where fluctuation-dissipation dynamics may be of relevance is the cosmological variation of fundamental constants driven by scalar fields. These could include e.g.~unknown scalars driving variations of the fine-structure constant, $\alpha$, or even the cosmological evolution of the SM Higgs field, which determines all fermion masses and in particular the electron-proton mass ratio $m_e/m_p$ (see e.g.~\cite{Calmet:2014qxa}). In the latter case dissipation could delay the electroweak phase transition, as we will discuss in this work for generic phase transitions, and potentially yield temporal variations of the electron-proton mass ratio $m_e/m_p$. Additionally, the associated fluctuations will also induce spatial variations of this ratio, which depending on their size and scale could in principle lead to observable effects.

This fluctuation-dissipation dynamics is not specific to near thermal equilibrium conditions. Whatever the statistical state is, a relation exists between the dissipation produced by the system and the fluctuations induced by the radiation bath. The near thermal equilibrium regime is, however, amenable to explicit calculations using well developed thermal field theory methods and will be the focus of this paper. The early Universe is generally believed to be in a near thermal equilibrium state and so these calculations based on thermal field theory have significant relevance to it. Nevertheless, there could be processes in the early cosmic stages where a scalar field  moves too quickly or the underlying microphysical processes are too slow to justify a near thermal equilibrium approximation. Thus it should be kept in mind that the calculations done in this paper could also be extended to these regimes; it would be a technical, albeit complicated, step further, but the underlying concept is the same as developed in this paper.

It is the goal of this work to set the stage for the study of cosmological fluctuation-dissipation dynamics within concrete particle physics models. We will discuss different examples of dissipation (and related noise) coefficients within the SM and beyond, exploring their distinct parametric regimes and domains of applicability. We will then outline some of the generic consequences of dissipation, particle production and induced fluctuations in the dynamics of cosmological scalar fields, both with and without the occurrence of phase transitions. We will focus on the post-inflationary dynamics, where the effects of fluctuation-dissipation dynamics remain largely unexplored, there existing already a considerable literature devoted to this topic in inflationary cosmology in the context of the above-mentioned warm inflation dynamics. 

To better illustrate the cosmological impact of these processes, we will analyze in detail two concrete scenarios. Firstly, we will consider a high-temperature phase transition in the early universe, where the associated Higgs field can dissipate its energy into fermionic modes through standard Yukawa interactions. In particular, we will show that, by slowing down the field's motion, dissipation will delay the phase transition, leading to additional entropy production and Hubble expansion that can dilute the abundance of dangerous relics such as topological defects. Furthermore, if the transition is sufficiently delayed, the Higgs field may come to dominate the energy balance and yield an additional (short) period of inflationary expansion. This results in a more efficient dilution of thermal relics, similarly to thermal inflation models, although dissipative friction can sustain accelerated expansion below the temperatures at which thermal effects can hold the Higgs field in the symmetric phase. Both thermal and dissipative (warm) inflation are, in fact, due to the same interactions between the Higgs field and the thermal bath degrees of freedom and may occur within the same cosmological phase transition, as we show in this work.

Secondly, we will consider the relaxation of a scalar field from a large post-inflationary value to its minimum energy configuration. We will show that its coupling to a B- or L-violating sector can lead to the dissipative production of a baryon or lepton asymmetry, respectively, in the spirit of the {\it warm baryogenesis} scenario proposed in \cite{BasteroGil:2011cx} in the inflationary context. To illustrate this generic mechanism, we develop a concrete model of {\it dissipative leptogenesis}, where dissipation results from the excitation and decay of heavy right-handed neutrinos, which gain a large Majorana mass from the coupling to a dynamical scalar field. As opposed to standard leptogenesis and other thermal baryogenesis scenarios, these are mainly produced off-shell, which allows for baryogenesis at parametrically low temperatures and therefore avoids the troublesome overproduction of thermal relics.

These two examples show that dissipative effects can have an important role in the cosmic history, particularly in addressing some of the most important puzzles in modern cosmology. We therefore hope that they motivate a more thorough exploration of this topic and of more general non-equilibrium processes in cosmology.

 This work is organized as follows. In Section \ref{FD} we show examples of how dissipation arises within common particle physics models, focusing in particular on the electroweak phase transition and Grand Unified Theories. In Section \ref{Phasetransitions} we describe the effects of fluctuation-dissipation dynamics in high-temperature phase transitions. In Section \ref{Leptogenesis} we describe the post-inflationary production of a baryon asymmetry through dissipative effects, describing in detail the {\it dissipative leptogenesis} scenario and its observational signatures. We summarize our main results and conclusions in Section \ref{Conclusion}, also discussing prospects for future research in this area.

%%%%%%%%%%%%%%%%%%%%%%%%%%%%%%%%%%%%%%%%%%%%%%%%%%%%%%%%%%%%%%%%%%%%%%%%%%%%%%%%%%%%%%%%%%%%%%%%%%%%%%%%%%%%%%%%%%%%%%%%%%%%%%%%%%%%%%%%
%%%%%%%%%%%%%%%%%%%%%%%%%%%%%%%%%%%%%%%%%%%%%%%%%%%%%%%%%%%%%%%%%%%%%%%%%%%%%%%%%%%%%%%%%%%%%%%%%%%%%%%%%%%%%%%%%%%%%%%%%%%%%%%%%%%%%%%%

\section{Fluctuation-dissipation dynamics in particle physics models}
\label{FD}

The several scalar fields employed in particle physics and associated cosmological models are typically not isolated systems and generically interact with other degrees of freedom. Their dynamics is therefore described by a quantum effective action that encodes the effects of all interactions with other fields. In the cosmological context, this effective action must take into account the non-trivial statistical states of both the dynamical field and the degrees of freedom with which it interacts. The black-body spectrum of the Cosmic Microwave Background and the successful predictions of BBN show that the universe was in a state very close to local thermal equilibrium for a great part of its early history, and we will henceforth assume that all relevant particle states always remain near this configuration.

For static fields the effective action reduces to a local effective potential, which takes the well-known Coleman-Weinberg form at leading order in a perturbative expansion \cite{Coleman:1973jx}. From the finite temperature effective potential one can derive the thermodynamic properties of the cosmological fields, such as their energy density, entropy and pressure, as well as thermal mass corrections. Static fields are, however, generically of little interest in cosmology, and for dynamical fields the effective action includes non-local effects beyond the leading effective potential approximation.

Time non-local effects may take different forms depending on the regime considered. The simplest case is the {\it adiabatic} regime, where the field varies on time scales that largely exceed the typical time scales of the relevant microphysical processes. This is for example the case of the inflaton field, which in the simplest scenarios is slowly rolling in order to produce a quasi-de Sitter phase. Local thermal equilibrium in the ambient heat bath can be maintained if scattering and/or decay processes within it are sufficiently fast, namely faster than Hubble expansion, such that an adiabatic approximation will be valid for large classes of cosmological scalar fields. 

In the adiabatic limit, a system has sufficient time to relax to an equilibrium configuration in response to the perturbing effect of the time non-local terms in the effective action, and linear response theory can be used to study the system's evolution. The leading time non-local effect is dissipation of the scalar field's energy into the degrees of freedom in the heat bath, which manifests itself through an effective friction term in the field's equation of motion. 

The simplest example of this is the creation and subsequent annihilation of particle-antiparticle pairs coupled to the background field, where this coupling makes the amplitude of creation and annihilation field-dependent. Suppose then that pairs are created at a time $t$ where the scalar field takes a value $\phi(t)$. They will then annihilate at time $t+ \delta t$ where the field has shifted by an amount $\delta\phi= \dot\phi \delta t +\dots$ in the adiabatic regime, and to leading order there will be a net particle production proportional to $\dot\phi$, resulting in a transfer of energy from the scalar field into the produced degrees of freedom. This will perturb the local thermal equilibrium in the ambient heat bath but the system can relax into a new equilibrium configuration if adiabatic dissipation is slower than other microphysical processes.

Dissipation corresponds to the systematic effect of the particles in the heat bath on the evolution of the field and the resulting friction opposes the latter's evolution through the creation and annihilation of particles in the heat bath in a field-dependent fashion, as outlined above. This is entirely analogous, for example, to the systematic friction force produced on a moving mirror by a rarefied gas of molecules that randomly hit the the mirror in a brownian motion \cite{0034-4885-29-1-306}. Much like this random brownian motion also results in an irregular motion of the mirror, fluctuations in the cosmological heat bath will also backreact on the evolution of the scalar field and introduce a degree of randomness. The two effects, fluctuations and dissipation, result from the same interactions between the scalar field and the heat bath and are thus interconnected. This is a general result that applies to large classes of dissipative systems in nature and is known as the {\it fluctuation-dissipation theorem}, the details of which depend on the statistical state of the system and its microscopic properties. 

The combined effect of fluctuations and dissipation leads to an effective Langevin-like equation for a cosmological scalar field interacting with the ambient heat bath of the form (see e.g. \cite{Morikawa:1986rp,Gleiser:1993ea,Berera:1996nv,Berera:1995wh,Berera:2007qm,Miyamoto:2013gna} and references therein):
\begin{equation} \label{langevin}
\ddot{\phi}+(3H+\Upsilon)\dot{\phi} -{1\over a^2}\nabla^2\phi+ V'(\phi) = \zeta~,
\end{equation}
where $\Upsilon$ denotes the dissipation coefficient and $\zeta$ the related random noise term, with the remaining terms yielding the usual Klein-Gordon equation in a flat FRW universe. In the adiabatic regime, the noise term is gaussian to leading order and its correlator satisfies the following fluctuation-dissipation relation in momentum-space \cite{Berera:2008ar,Ramos:2013nsa,Bastero-Gil:2014jsa}:
\begin{equation} \label{noise}
\langle \zeta(\mathbf{k},t) \zeta(\mathbf{k'},t')\rangle = \left[(1+2n(k))\frac{3 H^2\sqrt{1+ 4 \pi Q/3}}{\pi} + 2\Upsilon T\right]a^{-3}(2\pi)^3\delta^3(\mathbf{k} + \mathbf{k'})\delta(t - t')~,
\end{equation}
where we have also included a ``quantum noise'' contribution, given by the first term within brackets and $Q = \Upsilon/3H$. This results from a coarse-graining of the scalar field as employed in the stochastic approach to inflation \cite{Starobinsky:1986fx}, with short-wavelength (sub-horizon) field modes inducing an effective noise in the dynamics of the long-wavelength ``classical'' modes.  While the form in Eq.~(\ref{noise}) is obtained for a sharp mode splitting, a smooth filtering function generically results in a coloured noise distribution \cite{Winitzki:1999ve}. We have also included the effect of a generic phase-space mode distribution $n(k)$ \cite{Bastero-Gil:2014jsa}, which vanishes in the standard stochastic inflation approach, but becomes significant, in particular, when the scalar field is itself thermalized and $n(k)$ is the Bose-Einstein distribution. For example, for $T\gg H, |V''(\phi)|$ and $Q\ll1$, the first term within brackets becomes proportional to $HT$ for modes crossing the horizon.

As discussed above, the dissipative friction term is associated with a net particle creation in the ambient heat bath. One can then integrate Eq.~(\ref{langevin}) and average over the noise term to obtain the evolution of the scalar field's energy density, and use energy conservation to derive the associated equation for the heat bath: 
\begin{equation} \label{energy_conservation}
\dot{\rho_{\phi}} + 3H\dot{\phi}^2 = - \Upsilon \dot{\phi}^2, \qquad \dot{\rho_{\alpha}}+3H(\rho_{\alpha}+p_{\alpha}) = \Upsilon \dot{\phi}^2,
\end{equation}
where we take the heat bath to be described, to leading order, by a perfect fluid of density $\rho_\alpha$ and pressure $p_\alpha$. If it is composed of relativistic particles that thermalize sufficiently fast, the latter corresponds to a radiation fluid with equation of state $p_R= \rho_R/3$.

As we discuss below in more detail, we will be mainly interested in interactions between cosmological scalar fields and other (complex) scalar and fermionic degrees of freedom. Gauge interactions may also be of relevance for early universe cosmology, but since the main features of vector boson interactions are well described by scalar degrees of freedom we will not consider this case explicitly to simplify our discussion. We will thus consider a generic (renormalizable) Lagrangian of the form:
\begin{equation} \label{lagrangian_generic}
\mathcal{L}= f(\phi)|\chi|^2 + g\phi\bar{\psi}\psi~,
\end{equation}
where $\chi$ and $\psi$ denote complex scalar and fermion fields in the heat bath and $f(\phi)$ is a generic function of the dynamical scalar field we are interested in. 

\begin{figure}[htbp]
\includegraphics[scale=0.3]{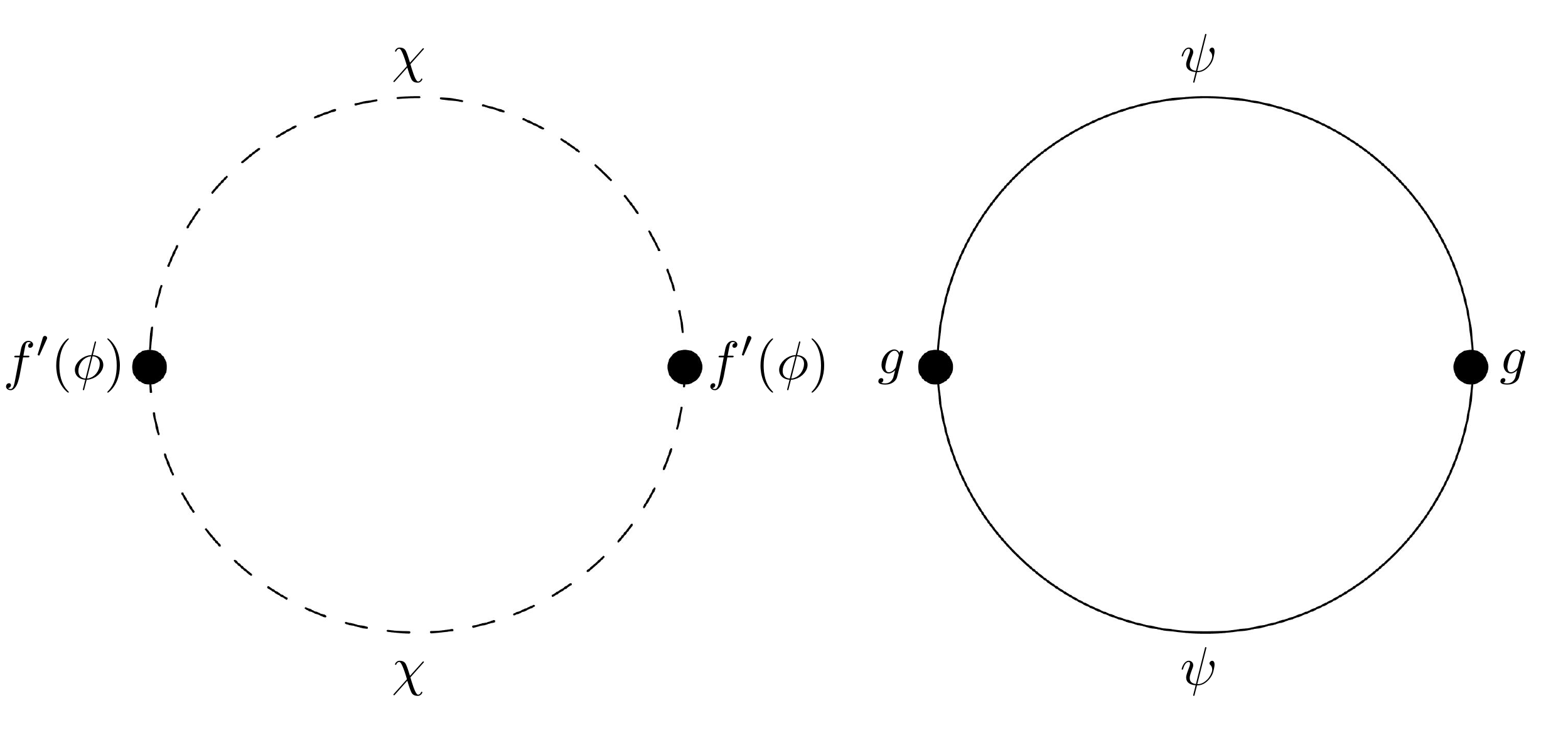}
\caption{Leading 1-loop contributions of scalar and fermion degrees of freedom to the dissipation coefficient. The thick dashed and solid lines indicate dressed propagators for scalars and fermions, respectively.}
\label{dissipation}
\end{figure}

The leading 1-loop contributions of these interactions to the effective action are illustrated in Fig.~\ref{dissipation} and, for a nearly-thermal heat bath at temperature $T=1/\beta$, these yield an adiabatic dissipation coefficient of the form:
\begin{equation} \label{dissipation_coeff_generic}
\Upsilon = \frac{f'(\phi)^2}{T} \int \frac{d^4p}{(2\pi)^4}\rho_{\chi}(p_0,{\bf{p}})^2n_B(p_0)(1+n_B(p_0)) + \frac{g^2}{2T} \int \frac{d^4p}{(2\pi)^4} \mathrm{Tr}\left[\rho_{\psi}(p_0,{\bf{p}})^2\right]n_F(p_0)(1-n_F(p_0))~,
\end{equation}
where $n_B(\omega) = (e^{\beta\omega}-1)^{-1}$ is the Bose-Einstein and $n_F(\omega) = (e^{\beta\omega}+1)^{-1}$ is the Fermi-Dirac distribution for particle modes of energy $\omega$. The functions  $\rho_{\chi}$ and $\rho_{\psi}$ represent the spectral functions of the scalar and fermion fields in the heat bath and can be computed from the corresponding (dressed) propagators at finite temperature, using e.g.~the real-time formalism \cite{Moss:2006gt,Berera:2008ar,BasteroGil:2010pb,BasteroGil:2012cm}. For example, in the scalar field case one obtains:
\begin{equation} \label{spectral_function}
\rho_\chi={4\omega_p \Gamma_\chi\over (p_0^2-\omega_p^2)^2 + 4\omega_p^2\Gamma_\chi^2}~,
\end{equation}
where $\omega_p=\sqrt{|\mathbf{p}|^2+m_\chi^2}$, with the field mass corresponding to its renormalized value including thermal corrections, and $\Gamma_\chi$ is the (finite temperature) decay width of the field. A similar, albeit more complicated expression, can be obtained in the fermionic case \cite{BasteroGil:2010pb}.

From Eq.~(\ref{dissipation_coeff_generic}) one can deduce a few generic aspects of dissipation coefficients in the adiabatic regime. Firstly, we see that if the fields $\chi$ and $\psi$ were in a trivial (vacuum) state, the dissipation coefficient would vanish. Dissipation is thus an effect intrinsic to the motion of the scalar field in the presence of a heat bath with non-trivial occupation numbers, corresponding as argued above to the systematic effect of the heat bath degrees of freedom on the field's motion. Secondly, the spectral functions correspond to the imaginary part of the field propagators and are consequently proportional to their decay width, as shown above. Hence, if the field's $\chi$ and $\psi$ were stable there would be no dissipation. However, at finite temperature interacting fields have always a non-zero decay width, arising from a combination of decays, inverse decays and Landau damping processes. Finally, the dissipation coefficient will in general be both field- and temperature-dependent. The temperature dependence is explicit in the distribution functions but will also arise in general in the masses and decay width of the fields. The field dependence can be explicit in the scalar case, for a generic function $f(\phi)$, but will also arise from the masses (and consequently the decay width of the fields), noting that from Eq.~(\ref{lagrangian_generic}) one obtains at tree-level $m_\chi^2=f(\phi)$ and $m_{\psi}^2=g^2\phi^2$. 

The integrals in Eq.~(\ref{dissipation_coeff_generic}) can be computed numerically in general, there existing, however, two approximate regimes where the computation can be performed analytically. In the {\it low temperature regime}, $m_\chi, m_{\psi_\chi}\gg T$, the distribution functions $n_B, n_F$ become exponentially (Boltzmann) suppressed for on-shell field modes, $p_0^2=\omega_p^2$, so that their contribution can be neglected. The main contribution in this case comes from off-shell or virtual modes with $|\mathbf{p}|,p_0\ll m_{\chi},m_{\psi}$, for which the spectral functions take a simple form, e.g. $\rho_\chi\simeq 4\Gamma_\chi/m_\chi^3$ in the scalar case. The {\it low-momentum} dissipation coefficient (LM) is thus approximately given by:
\begin{equation} \label{low-momentum}
\Upsilon^{LM} \simeq \frac{16 f'(\phi)^2}{T}\int\frac{d^4p}{(2\pi)^4}\frac{ \Gamma_{\chi}^2}{m_{\chi}^6} n_B(1+n_B) +
\frac{2 g^2}{T}\int\frac{d^4p}{(2\pi)^4}\frac{\mathrm{Tr}(\mathrm{Im}\Sigma^2)}{m_{\psi}^2}n_F(1-n_F)
\end{equation}
where $\mathrm{Im}\Sigma$ is the imaginary part of the fermion self energy, from which their decay width can be extracted in the conventional way. This approximation is valid in the narrow width limit where $m_i \gg \Gamma_i$ for $i=\chi, \psi$ (see \cite{BasteroGil:2012cm} for more details). The integrals involving the distribution functions and the decay widths can then be performed numerically \cite{BasteroGil:2010pb,BasteroGil:2012cm}. A simple example that we will consider below is the case where $f(\phi)\propto \phi^2$, with $m_\chi\propto \phi$ and  $\Gamma_\chi \propto m_\chi$. In this case, it is not difficult to see that $\Upsilon^{LM}\propto T^3/\phi^2$ for the scalar contribution, while the corresponding fermionic contribution is suppressed by further powers of $T/m_{\psi}\ll 1$ as shown in \cite{BasteroGil:2010pb}. 

In the opposite {\it high-temperature regime}, $m_\chi, m_\psi \ll T$, it is energetically possible to excite on-shell modes in the thermal bath and their occupation numbers are not Boltzmann-suppressed. These will then give the dominant contribution to the dissipation coefficient, and one can expand the spectral functions about their poles at $p_0=\omega_p$ to yield:
\begin{equation} \label{pole}
\Upsilon^{P} \simeq \frac{f'(\phi)^2}{T}\int\frac{d^3p}{(2\pi)^3} \frac{1} {\Gamma_{\chi}\omega_p^2} n_B(1+n_B) + \frac{2g^2}{T}\int\frac{d^3p}{(2\pi)^3}\frac{m_{\psi}^2}{\Gamma_{\psi}\omega_p^2} n_F(1-n_F)
\end{equation}
The 3-momentum integrals can then be easily computed analytically in different regimes (see e.g. \cite{BasteroGil:2012cm}). In particular, for light on-shell modes one typically obtains $\Gamma_i\propto m_i \propto T$, for $i=\chi,~\psi$, yielding $\Upsilon^P\propto \phi^2/T$ for scalar modes and $\Upsilon^P\propto T$ for fermionic modes \cite{BasteroGil:2010pb,BasteroGil:2012cm}.

In the general case, the dissipation coefficient receives contributions from both on-shell and off-shell modes, and numerical calculations show that adding both contributions yields a very good approximation to the full result. In particular, it has been observed that the on-shell contribution can be dominant for masses $m_i\gtrsim T$ despite the associated Boltzmann-suppression, particularly for small decay widths \cite{BasteroGil:2012cm}.

These two regimes will be relevant for different types of particle physics and cosmological scenarios. On the one hand, in a typical phase transition the relevant Higgs field is stabilized at the origin at high temperatures and starts rolling towards the minimum of its potential below a critical temperature (potentially after tunneling in a first order phase transition). In this case, the fields it couples to are initially light, and on-shell dissipation dominates. As $T$ decreases and the field value approaches the true minimum, these fields become heavier and the contribution of low-momentum modes will grow until it potentially dominates. On the other hand, a light scalar $\phi$ can attain very large values during inflation, after which it will eventually roll towards the minimum of its potential. In this case, off-shell modes will typically dominate initially, while on-shell modes will become increasingly more significant if $\phi$ evolves towards smaller values and $\chi,\psi$ become lighter.

In the following we give a series of examples, by no means exhaustive, of dissipation coefficients for dynamical scalar fields in the SM and its typically considered extensions, within different dynamical regimes that may be relevant for the cosmic history.

%%%%%%%%%%%%%%%%%%%%%%%%%%%%%%%%%%%%%%%%%%%%%%%%%%%%%%%%%

\subsection{Dissipation in the SM and supersymmetric extensions}

The SM gauge group, $SU(3)_c\times SU(2)_L\times U(1)_Y$, is broken spontaneously to $SU(3)_c\times U(1)_Q$ by the non-vanishing vacuum expectation value of the electroweak Higgs boson. In typical cosmological scenarios, the reheating temperature after inflation largely exceeds the critical temperature of the electroweak phase transition. Quarks, leptons and electroweak gauge bosons are relativistic and in thermal equilibrium, and their backreaction on the Higgs effective potential at high temperatures stabilizes the Higgs field at the symmetric minimum. As the universe cools down, the effective potential approaches its zero temperature form and the Higgs field will roll towards the finite vev that spontaneously breaks the electroweak symmetry. As the field rolls from the origin towards the broken minimum, we then expect dissipative processes to be mainly mediated by on-shell quarks and leptons, as well as the weak gauge bosons. The former, in particular, have the following well-known Yukawa couplings to the Higgs field:
\begin{equation} \label{SM}
\mathcal{L} \sim \lambda_e^{ij} \bar{e}_{R,i} \phi^\dagger  L_j + \lambda_u^{ij} \bar{u}_{R,i} \phi q_j + \lambda_d^{ij}\bar{d}_{R,i} \phi^{\dagger} q_j + \mathrm{h.c.}
\end{equation}
where we have suppressed weak isospin and color indices, while the indices $i,j$ label the fermion generations. At high temperatures, the decay width of quarks and leptons is given essentially by Landau damping terms from the above Yukawa interactions, as well as gauge interactions \cite{Satow:2010ia}. The relevant dissipation coefficient is thus of the on-shell form given in Eq.~(\ref{pole}). Including both Yukawa and gauge contributions to the fermions' decay width we obtain:
\begin{equation} \label{dissip_SM}
\Upsilon_{SM}^P \simeq \frac{288\zeta(3) T}{\pi^3}\sum_{i=1}^3 \left( \frac{(\lambda_e^{ii})^2}{(\lambda_e^{ii})^2 + 4(g_1^2 + 3 g_2^2)} + \frac{3(\lambda_u^{ii})^2}{(\lambda_u^{ii})^2 
+ 4(g_1^2+3g_2^2+8g_3^2)} +  \frac{3(\lambda_d^{ii})^2}{(\lambda_d^{ii})^2+ 4(g_1^2+3g_2^2+8g_3^2)}\right)~,
\end{equation}
where $g_i$ are the SM gauge couplings. We note that dissipation can also occur through the excitation of the $W^\pm$ and $Z$ gauge fields, although for simplicity we do not include this in the above expression. As $\phi$ increases and $T$ decreases eventually the masses of the SM particles will become heavier than the temperature and their on-shell contribution to dissipation becomes Boltzmann suppressed, with low-momentum dissipation of the form in Eq.~(\ref{low-momentum}) becoming the dominant contribution.

Another example based on the same symmetry group arises within the minimal supersymmetric extension of the SM (MSSM), where two Higgs doublets are required to break the electroweak symmetry and give masses to all quarks and leptons. The MSSM superpotential is given by:
\begin{equation} \label{MSSM}
W = \mu H_u H_d + y_{u}H_uQ U^c  + y_d H_d Q D^c + y_e H_d L E^c~.
\end{equation}
As in the SM, in this case one finds dissipative channels for both Higgs scalar components, $h_u$ and $h_d$, by exciting both fermion and sfermion degrees of freedom in the heat bath. At high temperatures these will have the forms $\Upsilon\propto \phi^2/T$ and $\Upsilon \propto T$ obtained above for sfermions and fermions, respectively, with $\phi= h_u, h_d$ (see also \cite{Kamada:2009hy}).

The simplest extension of the MSSM, known as the next-to-minimal supersymmetric SM (NMSSM), replaces the $\mu$-term in the superpotential by a trilinear term $g\Phi H_uH_d$, where $\Phi$ is a singlet chiral superfield. The effective $\mu$-term is then given by the vev of the scalar component of $\Phi$, a possibility that helps addressing the smallness of the {\it a priori} unconstrained $\mu$ parameter required for successful electroweak symmetry breaking (see e.g.~\cite{Ellwanger:2009dp}). One can then envisage scenarios where the singlet scalar field, $\phi$, is driven towards (or maintained at) a large value during inflation, after which it will roll towards $\langle \phi \rangle=\mu/g$. Its coupling to both Higgs scalar doublets $h_u$ and $h_d$ is of the form given in Eq.~(\ref{lagrangian_generic}) and explicitly:
\begin{equation} \label{NMSSM}
\mathcal{L}_S = g^2|\phi|^2(|h_d|^2 + |h_u|^2) + g\phi^\dagger h_d^{\dagger}\,y_u^{ij}\tilde{q}_i\tilde{u}_j^c + g\phi^\dagger\,  h_u^{\dagger}(y_d^{ij}\tilde{q}_i\tilde{d}_j^c + y_e^{ij}\, \tilde{l}_i\tilde{e}_j^c) +\mathrm{h.c.}~, 
\end{equation}
so that the singlet field can dissipate its energy through excitation of both scalar doublet components and their fermionic superpartners, which decay into the SM fermions and sfermions. If the initial field value is large, the dominant contribution to the dissipation coefficient is given by off-shell scalar modes as discussed above and the dissipation coefficient is approximately given by:
\begin{equation} \label{dissip_NMSSM}
\Upsilon_{NMSSM}^{LM} \simeq C_{\phi} \frac{T^3}{\phi^2}, \hspace{1cm} C_{\phi} \simeq \frac{1}{8\pi}\sum_{ij}\left(3(y_u^{ij})^2 +3(y_d^{ij})^2 + (y_e^{ij})^2\right)~,
\end{equation}
where $i,j$ run over family indices.

%%%%%%%%%%%%%%%%%%%%%%%%%%%%%%%%%%%%%%%%%%%%%%%%%%%%%%%%%

\subsection{Dissipation in Grand Unified Theories: an $SU(5)$ example}

There is significant evidence for the unification of the SM gauge couplings at high energy scales, particularly within the context of the MSSM \cite{Ellis:1990wk,Giunti:1991ta,Amaldi:1991cn,Langacker:1991an}, which points towards the existence of a larger gauge symmetry group. Several Grand Unified Theories (GUT) have been proposed where this gauge group is spontaneously broken into the SM gauge group through a Higgs-like mechanism, with $SU(5)$ and $SO(10)$ being the simplest and most studied examples [see e.g.~\cite{Ross:1985ai}]. In GUT models the relevant Higgs fields are coupled to gauge bosons and matter fields, such that fluctuation-dissipation dynamics may play an important role in their cosmological evolution.

If GUT symmetries are restored after inflation, for a sufficiently high reheating temperature, the Higgs fields roll from the symmetric point to the symmetry breaking minimum once the temperature drops below a critical value. This may be preceded by a tunneling event if the transition is first order, depending on the particle content of the GUT model \cite{Linde:1978px}, but dissipative rolling will always occur. This scenario may, however, be troublesome for cosmology since symmetry breaking typically leads to the generation of dangerous topological defects such as monopoles, which may overclose the universe.  Although post-inflationary symmetry restoration is appealing from the point of view of thermal GUT baryogenesis models, there are viable alternative mechanisms for the production of a cosmological baryon asymmetry such as the {\it dissipative baryo/leptogenesis} mechanism that we describe in Section IV. 

In the case where GUT symmetries are not restored during reheating, the relevant Higgs fields may nevertheless find themselves displaced from the symmetry breaking minimum after inflation. This occurs if the Higgs fields are light during inflation, being frozen at some initial value or even driven to larger values by random quantum fluctuations. Dissipation will then also be relevant in the post-inflationary eras as the fields roll towards the true minimum of their potential.

To illustrate the form of dissipative effects in GUT models, we consider the simplest case of $SU(5)$, bearing in mind that similar processes will generically occur for higher-rank gauge groups where many other dynamical scalars and dissipative channels may be present. In fact, in Section IV we will consider the particular example of a scalar field responsible for the Majorana mass of right-handed neutrinos and which is naturally embedded in $SO(10)$ models.

$SU(5)$ is broken into the SM gauge group by the vev of an adjoint Higgs field, $24_H$, which gives masses to the gauge and fundamental Higgs field components that are associated with the broken symmetries. The adjoint scalar potential takes the form:
\begin{equation} \label{adjoint_potential}
V(24_H) = -\mu^2 \mathrm{Tr}[ 24_H^2] + a \mathrm{Tr}[24_H^2]^2 + b \mathrm{Tr}[ 24_H^4] + c \mathrm{Tr}[24_H^3]~, 
\end{equation}
which, in certain parametric regimes, has an absolute minimum in the direction $\phi\,\text{diag}( 2,2,2,-3,-3)/\sqrt{30}$ that preserves the SM gauge group. Interactions between the adjoint and fundamental Higgs fields are given by:
\begin{eqnarray} \label{adjoint_fundamental_int}
\mathcal{L}_{s} = -\frac{A^2}{2}5_H^{\dagger}5_H+\frac{B}{4}(5_H^{\dagger}5_H)^2 + C\, 5_H^{\dagger}5_H \mathrm{Tr}[24_H^2] +D\, 5_H^{\dagger} 24_H^2 5_H + E\, 5_H^{\dagger}24_H 5_H~,
\end{eqnarray}
while the latter is coupled to the SM matter fermions in the $10$ and $\bar{5}$ representations via Yukawa couplings of the form:
\begin{eqnarray} \label{yukawas_su5}
\mathcal{L}_{Y} = Y_5^{ij}\bar{5}_{F\,i}10_{F\,j}5_H^* + \frac{1}{8}\epsilon_5Y_{10}^{ij}10_{F\,i}10_{F\,j}5_H+\mathrm{h.c.}
\end{eqnarray}
Decomposing these fields in terms of SM representations we find the following interaction Lagrangian involving the symmetry breaking scalar direction $\phi$, the doublet and triplet Higgs fields, $H$ and $T$, and the SM quarks and leptons:
\begin{eqnarray} \label{int_su5}
\mathcal{L}_{int} &=& \left(-\frac{A^2}{2}+\frac{2E}{\sqrt{30}}\phi+\frac{2D}{15}\phi^2+C\phi^2\right) |T|^2 + \left(-\frac{A^2}{2}-\frac{3E}{\sqrt{30}}\phi+\frac{3D}{10}\phi^2 +C\phi^2\right)|H|^2 \nonumber \\
&-&Y_{5}^{i j}(l_i q_j + d^c_iu^c_j)T^* - Y_{10}^{i j}\left(\frac{1}{2}q_iq_j + u^c_ie^c_j\right)T
+Y_5^{ij} (L_ie_j^c + q_i d_j^c)H^* - Y_{10}^{ij}q_i u_j^c H + \mathrm{h.c.}   \label{SU(5)}
\end{eqnarray}
As discussed above there are also gauge interactions, but for illustrative purposes we will restrict ourselves to dissipative effects associated with the scalar and Yukawa interactions given above. Let us consider, in particular, the case where $\phi$ has a large vev after inflation that is displaced from its true minimum. The leading $C$-terms in Eq.~(\ref{int_su5}) give a large mass to the Higgs doublet and triplet, which are initially equal due to the large field vev, while a doublet-triplet mass splitting will only arise close to the minimum. These terms are of the generic form given in Eq.~(\ref{lagrangian_generic}) for $\chi=H, T$ and we expect the main contribution to dissipation in this regime to correspond to virtual Higgs modes decaying into quarks and leptons. From Eq.~(\ref{low-momentum}) we then obtain the following dissipation coefficient:
\begin{equation} \label{dissip_su5}
\Upsilon^{LM}_{SU(5)} \simeq {0.44\over C^2} {T^7\over\phi^6}\sum_{i,j}\left[ 10(Y_5^{ij})^4 +8(Y_{10}^{ij})^4\right] ~.
\end{equation} 
In a supersymmetric realization of $SU(5)$ the SM fermions have scalar superpartners and, due to the holomorphic nature of the superpotential, two distinct Higgs fields in the $5$ and $\bar{5}$ representation are required. The relevant part of the superpotential is given by:
\begin{equation} \label{susy_su5}
W =g\bar{5}_H 24_H 5_H + M \bar{5}_H 5_H + Y_5^{ij} \bar{5}_{F\,i} 10_{F\,j}\bar{5}_H +\frac{1}{8}\epsilon_5 Y_{10}^{ij}10_{F\,i} 10_{F\,j} 5_H~,
\end{equation}
where the relevant scalar interactions are:
\begin{eqnarray} \label{susy_su5_int}
\mathcal{L}_s &=& Y_5^{ij}\left({2g\phi\over\sqrt{30}}+M\right)\tilde{t_u}^{\dagger}
(\tilde{d}^c_i\tilde{u}^c_j-\tilde{l}_i\tilde{q}_j) + Y_{10}^{ij}\left({2g\phi\over\sqrt{30}}+M\right)\tilde{t_d}^{\dagger}
\left(\tilde{e}^c_i\tilde{u}^c_j+{1\over2}\tilde{q}_i\tilde{q}_j\right)  \nonumber\\
&+& Y_{10}^{ij}\left(M-{3g\phi\over\sqrt{30}}\right)\tilde{h}_d^\dagger\tilde{q}_i\tilde{u}_j^c + Y_5^{ij}\left(M-{3g\phi\over\sqrt{30}}\right)\tilde{h}_u^\dagger(\tilde{l}_i\tilde{e}^c_j+\tilde{q}_i\tilde{d}^c_j) +\mathrm{h.c.}
\end{eqnarray}
As in the non-SUSY model, the low-temperature regime for dissipation will be the relevant one after inflation if $\phi$ attains a large vev, $g\phi \gg M$, and the GUT symmetry is not restored. Dissipation is in this case dominantly mediated by virtual scalar doublet and triplet Higgs modes that decay mainly into sfermion fields, as shown in \cite{BasteroGil:2010pb} for generic SUSY models of this form. The dissipation coefficient is then given by:
\begin{equation} \label{dissip_su5_susy}
\Upsilon^{LM}_{SSU(5)_{}} \simeq \frac{1}{16\pi}\frac{T^3}{\phi^2}\sum_{i,j}\left[ 10(Y_5^{ij})^2 +8(Y_{10}^{ij})^2\right]~.
\end{equation}
Note that dissipative effects will be more pronounced in this case compared to the non-SUSY model, since the dissipation coefficient is less suppressed by powers of $T/m_{\chi}$, where $\chi$ generically denotes the doublet and triplet Higgs scalars involved.

%%%%%%%%%%%%%%%%%%%%%%%%%%%%%%%%%%%%%%%%%%%%%%%%%%%%%%%%%%%%%%%%%%%%%%%%%%%%%%%%%%%%%%%%%%%%%%%%%%%%%%%%%%%%%%%%%%%%%%%%%%%%%%%%%%%%%%%%%%%%%%%%%%%%%%%%%%%%%%%%%%%%%%%%%%%%%%%%%%%%%%%%%%%%%%%%%%%%%%%%%%%%%%%%%%%%%%%%%%%%%%%%%%%%%%%%%%%%%%%%%%%%%%%%%%%%%%%%%%%%%%%%%%%%%%%%%%%%%%%%%%%%

\section{Fluctuation - dissipation dynamics in cosmological phase transitions}
\label{Phasetransitions}

Despite the numerous studies in the context of condensed matter systems, the dynamics of phase transitions in fundamental particle physics and cosmology remains largely unexplored. The recent discovery of the Higgs boson at the LHC, so far consistent with the SM predictions for the spontaneous breaking of the electroweak gauge symmetry, is the first experimental hint for the occurrence of a fundamental phase transition in the cosmic history and motivates further exploration of this topic. Moreover, the apparent unification of gauge couplings suggests, as discussed above, that one or more phase transitions may have occurred in the early stages of the universe's history, spontaneously breaking a higher-ranked gauge group into $SU(3)_c\times SU(2)_L\times U(1)_Y$, in potentially several stages of a progressively lower degree of symmetry.

Cosmological Higgs fields are coupled to matter fields and gauge bosons, and the effects of dissipation and associated fluctuations will necessarily play a role in the evolution of the fields from a symmetric to a spontaneously broken symmetry phase. In this section, we will discuss several potential effects of fluctuation-dissipation dynamics in generic cosmological phase transitions.

%%%%%%%%%%%%%%%%%%%%%%%%%%%%%%%%%%%%%%%%%%%%%%%%%%%%%%%%%%%%%%%%%%%%%%%

\subsection{Thermal fluctuations and topological defects}

The cosmological evolution of a generic Higgs field in the process of spontaneous symmetry breaking follows a Langevin-like equation of the form (\ref{langevin}), with both the noise term on the right-hand side and the dissipative friction term on the left-handed side playing an important role at different dynamical stages. Fluctuations will be primarily significant at the onset of the phase transition, just below the critical temperature at which the symmetric Higgs value can no longer be stabilized by thermal effects. In particular, in the absence of random fluctuations the field would remain at the unstable symmetric minimum, since this is nevertheless an extremum of the effective potential. The noise term in the Langevin-like equation is thus crucial in inducing the phase transition and in determining the direction within the vacuum manifold towards which the field's evolution will proceed.

Since the Higgs field is, on average, at rest at the onset of the phase transition, its dynamics will be initially governed by the gaussian and white noise term in the adiabatic regime. As discussed earlier, the stochastic noise term encodes the effective backreaction of the ambient heat bath, also incorporating the inherent quantum nature of the field. On the one hand, the backreaction of the heat bath is directly related to the dissipation coefficient through the fluctuation-dissipation theorem and given by the term proportional to $\Upsilon$ in the noise correlator (\ref{noise}). On the other hand, the remaining quantum noise term, proportional to the Hubble parameter $H$, can be deduced from a coarse-graining of the quantum Higgs field, with short wavelength modes that are well within the Hubble horizon backreacting on the longer wavelength modes that one is interested in following. This stochastic approach has  e.g.~been successful in describing field fluctuations in both warm and cold inflation regimes \cite{Ramos:2013nsa}.

The phase transition will then initially be driven by the quantum and thermal/dissipative noise terms, which randomly kick the Higgs field away from and towards the symmetric minimum. This will proceed until the amplitude of the noise term becomes sub-dominant compared to the ``classical" terms in the equation of motion, i.e.~roughly when $\sqrt{\Upsilon T} \lesssim V'(\phi)/H^2$ for strong dissipation. Random fluctuations will then effectively cease and the subsequent field dynamics will essentially be classical. However, the field is now spatially inhomogeneous and the classical evolution will drive it to different directions in the vacuum manifold at distinct spacetime points. The classical dynamics can nevertheless homogenize the field within causally connected patches, determined by the field's correlation length, $\xi_c$, that is at most the size of the cosmological horizon.  If the Higgs field is relativistic at this stage, $\xi_c\sim 1/T\ll H^{-1}$, and so the temperature at which the noise term becomes inefficient will set the size and consequently the abundance of any topological defects that may form once the field settles into the lowest energy configuration. Some preliminary studies for the formation of topological defects in phase transitions including the effects of both thermal noise and dissipation have been performed in \cite{Laguna:1997cf, Stephens:1998sm,PhysRevD.52.3298}. It would be interesting to further explore this in the context of the concrete particle physics models discussed above and within realistic cosmological settings. This is, however, beyond the scope of the present work, where we will focus on the dissipative classical evolution.

One other related consequence of the noise term should nevertheless be pointed out. The inhomogeneity of the Higgs field resulting from its initial random motion will also induce a spatial variation of its gauge quantum numbers, generically sourcing magnetic fields \cite{Vachaspati:1991nm, Grasso:2000wj}. Their strength will then also be determined by the correlation length at the time when the noise term becomes inefficient and, if sufficiently large, this may sow the seeds for galactic magnetic fields.

Although here we will not pursue these issues in further detail, it is worth emphasizing that the Langevin-like equation (\ref{langevin}) gives a fundamental framework for these studies. Given a particle physics model, one can compute the dissipation coefficient and associated noise term from first principles, as explicitly done in the previous section for several examples, and use this equation to determine both the quantum and classical dynamics. This allows one to determine the correlation length, the density of topological defects or the strength of generated magnetic fields in a rigorous way. In this way, there is no need to simply employ statistical arguments to derive these quantities and the field evolution can be completely determined for arbitrary initial conditions.

%%%%%%%%%%%%%%%%%%%%%%%%%%%%%%%%%%%%%%%%%%%%%%%%%%%%%%%%%%%%%%%%%%%%%%%%%%%%%%%%%%%%%%%%%%%%%%%%%%%%%%%%%%%%%%%%%%%%%%%%%%%%%%%%%%%%%%%%

\subsection{Dissipative effects: entropy production and additional inflation}

Once the effects of the thermal and quantum noises become sub-dominant, the field's evolution becomes classical and is driven by the competition between the scalar potential's slope and the effects of dissipative and Hubble friction. To analyze the concrete effects of dissipation, which have so far been overlooked in the literature and, as we will show, may play an important role, we will consider a generic toy model where a real Higgs field is coupled to fermions through standard Yukawa couplings of the form in Eq.~(\ref{lagrangian_generic}). This can be easily extended to concrete particle physics models such as the electroweak phase transition or GUT phase transitions by considering the appropriate couplings, particle content and the properties of the vacuum manifold, as illustrated in the previous section.

The fermions induce, as discussed before, both local and non-local corrections to the effective action of the Higgs field. The leading effect of the former are finite temperature corrections to the effective potential, with zero-temperature corrections playing a sub-dominant role that we will for simplicity discard in our analysis. Thermal corrections are significant for relativistic fermions, $m_\psi\ll T$, namely inducing a thermal mass for the Higgs field, while for $m_\psi\gtrsim T$ these corrections are Boltzmann-suppressed and thus irrelevant to the dynamics. The general form of the thermal mass can be obtained by numerical integration, but for our purposes it is sufficiently accurate to explicitly multiply the high-temperature result by a Boltzmann factor, yielding for the effective Higgs potential:
\begin{equation} \label{Higgs_potential}
V(\phi, T) = \frac{\lambda^2}{4}(\phi^2-v^2)^2 + {1\over2}\alpha^2T^2\phi^2\exp\left(-\frac{m_{\psi,T}}{T}\right)~,
\end{equation}
where $\alpha^2=g^2N_F/6$, with $N_F$ denoting the number of Dirac fermion species. In this expression, the first term corresponds to the simplest symmetry-breaking potential with minima at $\phi=\pm v$ and the second term is the leading thermal correction. In the fermion mass $m_{\psi,T}^2=g^2\phi^2+h^2T^2$ we also include thermal corrections from coupling to different species in the heat bath, including e.g.~gauge fields, and which we generically parametrize with an effective coupling $h\lesssim 1$. At high temperatures, the second term dominates the effective potential and the Higgs field is thus stabilized at the origin, with $m_{\psi,T}\simeq h T\lesssim  T$, while at low temperatures it will roll towards one of the minima at $\phi=\pm v$. The symmetric minimum becomes unstable at a critical temperature:
\begin{equation} \label{critical_temp}
T_c= {\lambda v\over \alpha}.
\end{equation}
As soon as the field begins rolling towards the symmetry breaking minimum, it will feel the friction effect of the fermion heat bath. The relevant dissipation coefficient corresponds to the on-shell excitation of light fermions with a thermal decay width and was first computed in \cite{Yokoyama:1998ju}, yielding:
\begin{equation} \label{dissip_YL}
\Upsilon \simeq 11.2 N_F T \exp\left(-\frac{m_{\psi,T}}{T}\right)~.
\end{equation}
Note that, as above, we have multiplied the high-temperature result by a Boltzmann factor which will cut-off on-shell dissipation at low temperatures, $T\lesssim g\phi$, which is sufficiently accurate for our purposes. As we have seen above, virtual modes will also induce dissipation in the latter regime, but since this is a significantly smaller effect we will discard it in our analysis to a first approximation. Also notice that the SM high-temperature dissipation coefficient given in Eq.~(\ref{dissip_SM}) coincides with this expression if one discards gauge interactions. The thermal width of the fermions is roughly given by $\Gamma_\psi\sim m_{\psi,T}^2/T \lesssim T$, and adiabaticity of the dissipative process requires $\Gamma_\psi\gtrsim H$, which is easily satisfied in the radiation-dominated era, where $H\simeq \sqrt{\pi^2/90} g_*^{1/2} T^2/M_P \ll T$. For simplicity, we will assume that the radiation bath is made exclusively of fermions, taking $g_* = 7N_F/4$, although one can easily include other relativistic degrees of freedom.

With the form of the effective potential and dissipation coefficient, we may thus describe the dynamics of the phase transition by solving the system of coupled Higgs-radiation equations, given by:
\begin{equation}\label{DE}
\ddot{\phi} + (3H+\Upsilon)\dot{\phi} + V'(\phi) \simeq 0~, \qquad
\dot{\rho}_R + 4H\rho_R = \Upsilon \dot{\phi}^2~,
\end{equation}
where as discussed above we neglect the effects of the noise term, and $\rho_R=(\pi^2/30) g_*T^4$ assuming a nearly-thermal equilibrium state. Dissipation thus plays two distinct roles in the dynamics, on the one hand damping the field's motion and, on the other hand, sourcing the radiation bath through the production of fermion modes.

If the energy density in the scalar field is sufficiently large, it may come to dominate the energy density before the universe cools down to below the critical temperature. This occurs for $g^2N_F\gtrsim 2\pi \sqrt{g_*}\lambda$, thus inducing a period of {\it thermal inflation} along the lines proposed in  \cite{Lyth:1995hj} and which may help diluting the abundance of dangerous thermal relics produced e.g.~during reheating or earlier cosmological phase transitions. This additional period of inflation can typically last only for a few e-folds until the critical temperature is reached, with the field then rolling towards the symmetry-breaking minimum and oscillating about it.

In the presence of dissipation, the dynamics can be quite different below the critical temperature and an interesting alternative/addition to thermal inflation arises. Firstly, we note that in the radiation era $\Upsilon \gtrsim T \gg H$ and so the main source of friction is dissipation into fermionic modes rather than Hubble expansion. The field's motion will then be overdamped for $\Upsilon\gtrsim |m_\phi| $, where $m_\phi^2=V''(\phi)$, and underdamped otherwise. For relativistic fermions, close to the origin we have $m_\phi^2\simeq  \alpha^2(T^2-T_c^2)$, while for $\phi=\pm v$ the field mass is $m_\phi^2=\alpha^2(2T_c^2+T^2)$. This means that the field's trajectory from the symmetric to the symmetry-breaking minimum will be overdamped if during the transition the temperature is above $0.05 g T_c/\sqrt{N_F}$, which is parametrically below the critical value. This implies that instead of oscillating about $\phi=\pm v$, the field will smoothly evolve towards this value. When the motion is overdamped, the scalar field equation reduces to a slow-roll equation of the form:
\begin{equation}\label{slow-roll}
3H(1+Q)\dot\phi \simeq -V'(\phi)~,
\end{equation}
where $Q\equiv \Upsilon/3H$, such that in the radiation era $Q\sim 8.5 \sqrt{N_F}M_P/T\gg 1$ as argued above and the field's evolution occurs in a strong dissipation regime. Furthermore, the field may remain close to the origin and mimic a cosmological constant if its energy density does not vary significantly within a Hubble time:
\begin{equation}\label{slow-roll_cond}
{1\over H}{\dot{\rho}_\phi\over \rho_\phi}\simeq  -3(1+Q){\dot\phi^2\over V}\simeq -2{\Omega_\phi \epsilon_\phi\over 1+Q}\ll 1~,
\end{equation}
where $\epsilon_\phi= (M_P^2/ 2)(V'(\phi)/V(\phi))^2$ is the slow-roll parameter typically considered in slow-roll inflationary models and $\Omega_\phi=\rho_\phi/\rho_{total}$ is the Higgs field relative abundance. This condition essentially ensures that the Higgs field does not dissipate a significant fraction of its energy density into the heat bath on cosmological time scales, thus sustaining a cosmological constant-like behavior for small kinetic energy. Note also that the condition above reduces to the slow-roll condition $\epsilon_\phi\ll1 $ in non-dissipative (cold) inflationary models, where the scalar field is the dominant component, and to the slow-roll condition $\epsilon_\phi\ll 1+Q$ in dissipative (warm) inflation scenarios. It moreover shows that a constant energy density is easier to maintain when the field is subdominant, $\Omega_\phi<1$.

\begin{figure}[htbp]
\centering
\includegraphics[scale=0.4]{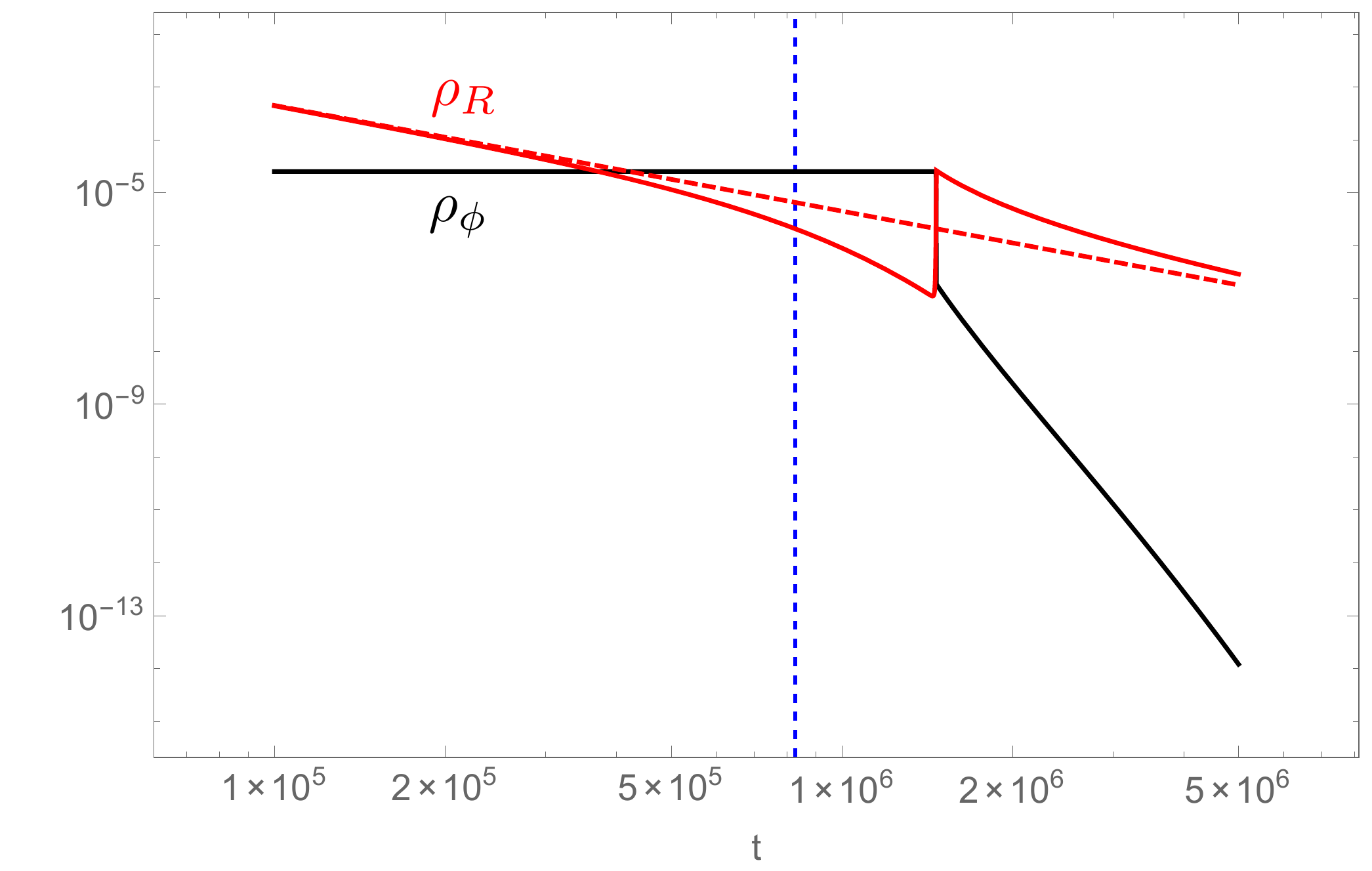} 
\caption{Evolution of the field and radiation energy density during a phase transition (solid lines), for a case where the field comes to dominate the energy density. The dashed red line shows the evolution of the radiation energy density in the absence of a phase transition. In this example, $\lambda = 0.01$, $N_F=10$, $v=10^{15}$GeV, $g^2=1/N_F$ and $h=0.1$. Time here is shown in units of the scale $10^{15}$ GeV. The vertical dashed line indicates the time when the critical temperature is reached.}
\label{domination}
\end{figure}

Close to the origin we find $\epsilon_\phi\simeq 8 (M_P/v)^2(\phi/v)^2$, which can be small if the field is very close to the origin. Inflation is, however, hard to maintain in the absence of dissipation with this type of ``hill-top" potential since the curvature parameter $\eta_\phi = M_P^2 V''/V\simeq -4 M_P^2/v^2$ is too large unless $v\gtrsim M_P$. Dissipation into the heat bath alleviates this constraint by overdamping the field's motion as shown above. As first shown in \cite{Yokoyama:1998ju, Berera:1998px}, it is hard to obtain a very long period of inflation with the dissipation coefficient in Eq.~(\ref{dissip_YL}), in particular the 50-60 e-folds required to solve the horizon and flatness problems, since the fermion mass increases as the field moves towards the minimum and eventually dissipation becomes Boltzmann-suppressed. We note, however, that for supersymmetric models in the low-temperature regime, where dissipation is dominantly mediated by low-momentum scalar field modes, fully successful models of warm inflation have been developed (see e.g. \cite{BasteroGil:2012cm, BasteroGil:2009ec, BasteroGil:2010pb,Bartrum:2013fia}). In such scenarios dissipation can sustain both the slow-roll dynamics of the inflaton field and the temperature of the radiation bath for a sufficiently long period. Nevertheless, the dissipation coefficient in Eq.~(\ref{dissip_YL}) can sufficiently overdamp the field's motion to allow for a few e-folds of inflation which, analogously to thermal inflation, can dilute dangerous relics generated prior to the phase transition. The slow-roll equation (\ref{slow-roll}) can be solved when the field is close to the origin in a radiation-dominated universe, yielding:
\begin{equation}\label{slow-roll_sol}
\phi\propto \exp\left[{0.1 \alpha^2\over N_F^{3/2}}{M_P\over T}\left(\left({T_c\over T}\right)^2-3\right)\right]~,
\end{equation}
so that the field increases exponentially below the critical temperature. Since $\epsilon_\phi \Omega_\phi /Q \propto \phi^2/T^3$ in this case, slow-roll can only be maintained for a finite period of time. The Higgs field may become dominant at a temperature $T_\phi\simeq g_*^{-1/4}\alpha T_c/\sqrt{\lambda}$, so if the scalar self-coupling is sufficiently small the field will overcome the radiation before the end of the slow-roll regime and induce a period of inflation. In Figure \ref{domination} we show a numerical example of a phase transition close to the GUT scale where a short period of inflationary expansion occurs. 

It is clear in this figure that the field's motion is always overdamped, exhibiting no field oscillation, even though slow-roll is only maintained for a finite period, with inflation lasting in this example for $\sim 1.5$ e-folds. The field then evolves quickly to the symmetry-breaking minimum, which is actually time-dependent until the field's thermal mass becomes exponentially suppressed. We note that this transition is fast in terms of the cosmological Hubble time, although still adiabatic from the microphysical perspective. We emphasize that dissipation prevents the field from oscillating about the minimum, as opposed to what is commonly considered in phase transitions when this effect is not taken into account. Therefore, in the presence of dissipation, the Higgs field will not behave as pressureless matter after the phase transition.

During the Higgs-dominated phase, the radiation density is diluted exponentially by the accelerated expansion until the end of the slow-roll regime. With the increase in the field's velocity, the dissipative source term in the radiation evolution equation (\ref{DE}) grows substantially, allowing radiation to once more become the dominant component. We note that in warm inflation models where 50-60 e-folds of inflationary expansion can be sustained, generically in the low-$T$ rather than the high-$T$ regime considered here, $\rho_R$ typically reaches a quasi-steady evolution with the dissipative source term balancing the Hubble dilution effect. In the example shown above, inflation does not last sufficiently long for this quasi-equilibrium to be reached, with first dilution and then dissipation playing a dominant role in the radiation evolution.

\begin{figure}[h!]
\centering
\includegraphics[scale=0.4]{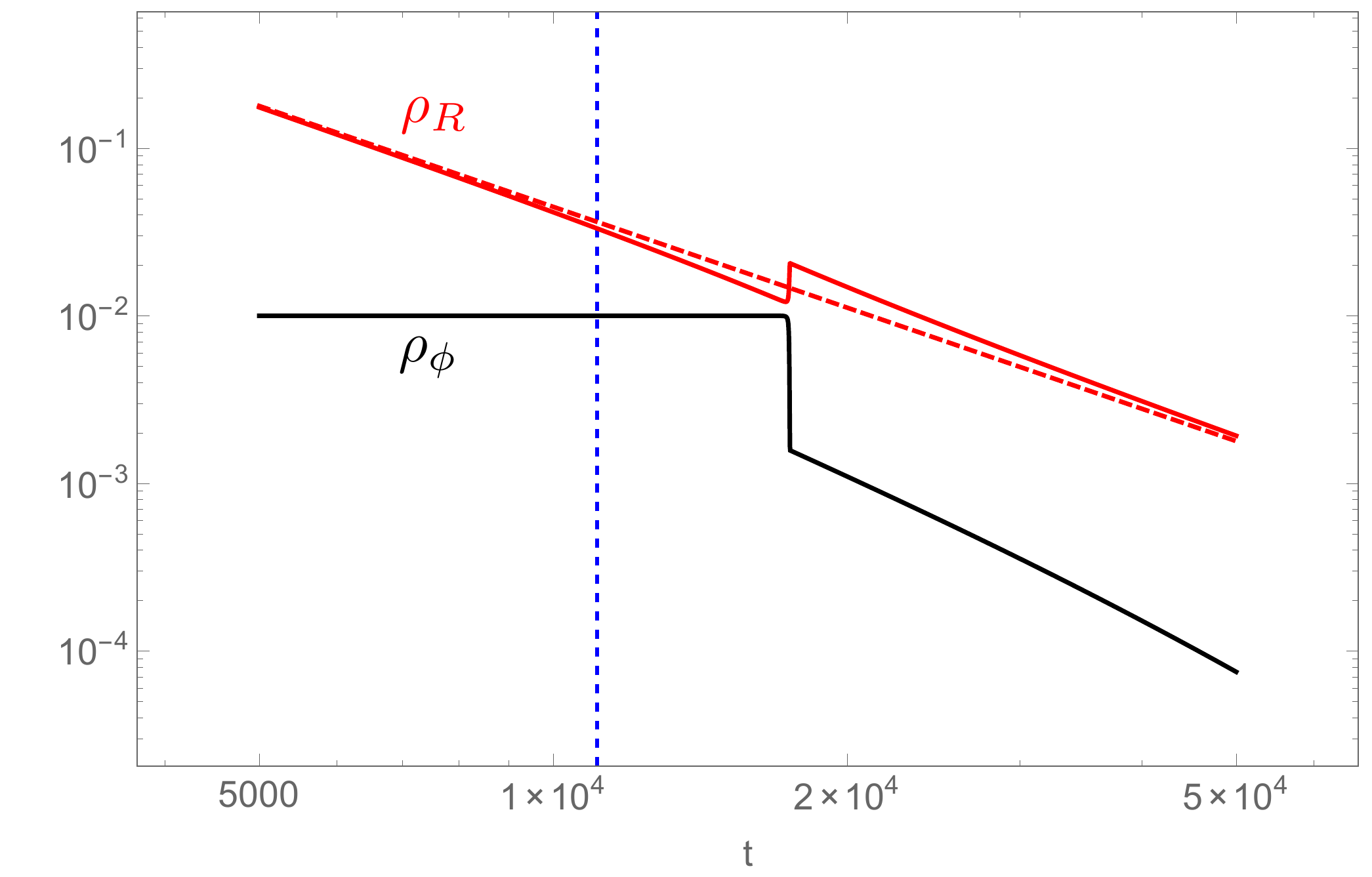} 
\caption{Evolution of the field and radiation energy density during a phase transition (solid lines), for a case where the field never dominates the energy density. The dashed red line shows the evolution of the radiation energy density in the absence of a phase transition. In this example, $\lambda = 0.2 $, $N_F = 1$, $v=10^{15}$ GeV, $g^2=1/N_F$ and $h=0.1$. Time here is shown in units of the scale $10^{15}$ GeV. The vertical dashed line indicates the time when the critical temperature is reached.}
\label{no-domination}
\end{figure}

A crucial point to emphasize is that the expansion history can be significantly modified even if inflation does not occur, i.e.~in parametric regimes where slow-roll cannot be sustained until the field can dominate. On the one hand, when the field is slowly rolling, its energy density increases the expansion rate, therefore diluting the ambient radiation more quickly even if it is sub-dominant. On the other hand, once slow-roll is over and the field quickly settles into the symmetry-breaking minimum, it can dissipate a significant part of its energy density into the heat bath. This is illustrated in Figure \ref{no-domination}, where slow-roll ends just before the scalar field's abundance becomes comparable to the heat bath.

\begin{figure}[h!]
\centering
\includegraphics[scale=0.4]{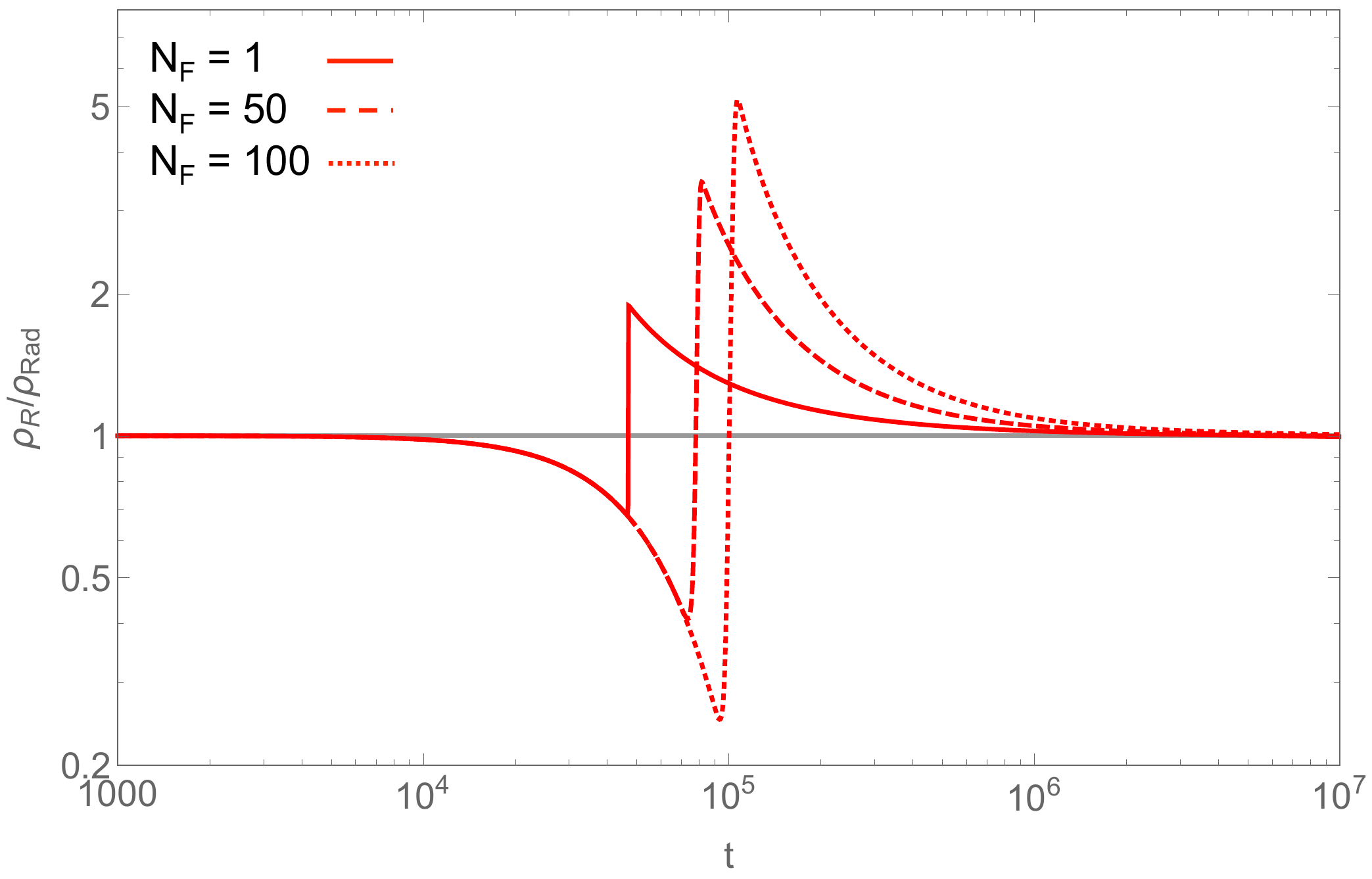} \hspace{0.2cm}
\includegraphics[scale=0.4]{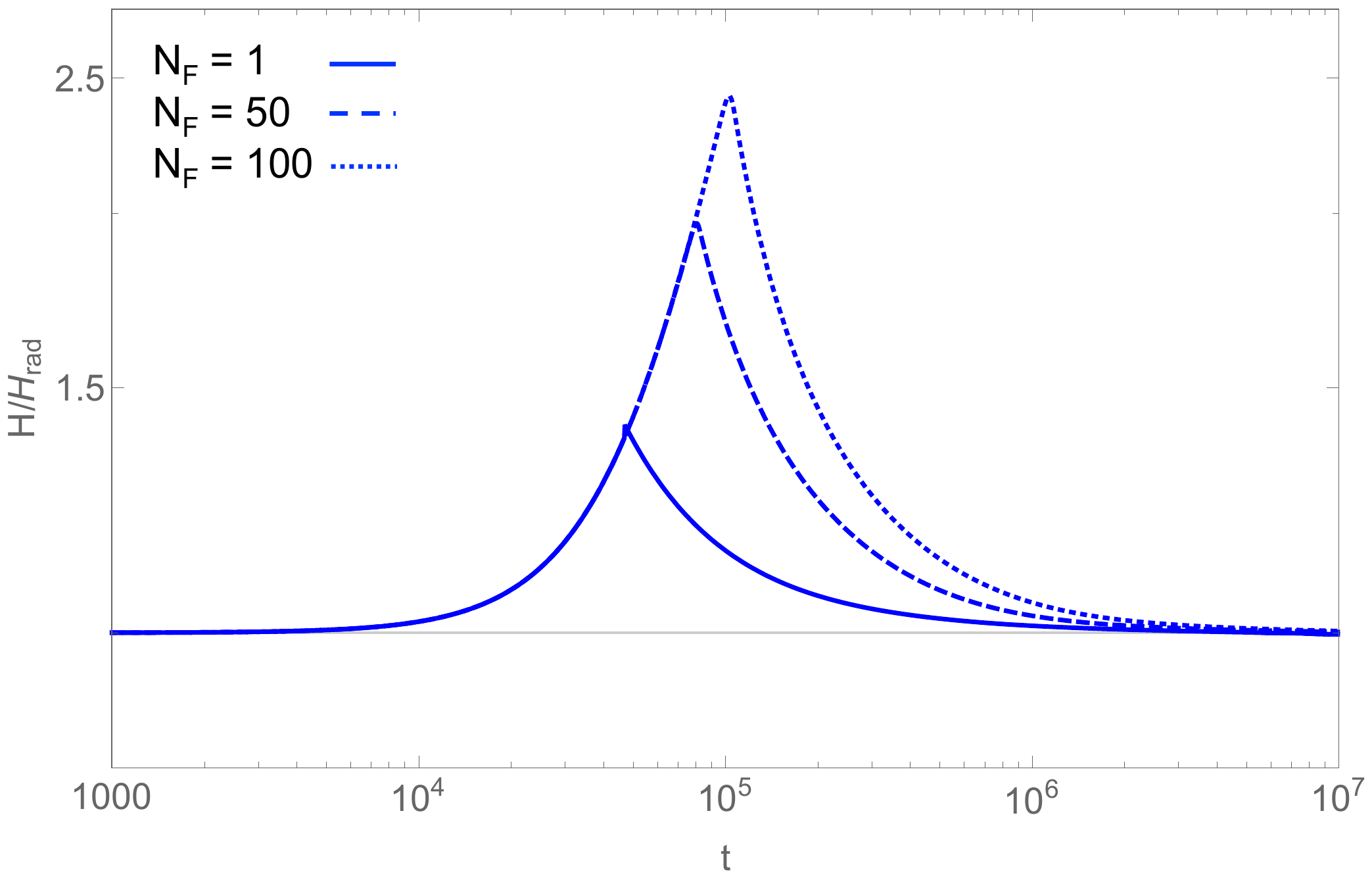}
\caption{In the left (right) hand plot we show the evolution of the radiation energy density (Hubble parameter) compared to that of a standard radiation dominated universe. In this example, $\lambda=0.1$, $v=10^{15}$ GeV, $g^2=1/N_F$ and $h^2 = 0.1$, Time here is shown in units of the scale, $10^{15}$ GeV.}
\label{naexpansion}
\end{figure}

This figure clearly shows that the most significant effects occur at the end of the slow-roll regime, when the Higgs field's relative abundance is maximal, leading first to a dilution and then to an increase in the radiation energy density. The latter eventually relaxes to the value it would have in the absence of a phase transition, since the relation $\rho_R(t)$ is an attractor of the Friedmann equation in a radiation-dominated universe. One can solve the equation of motion for radiation in the absence of a dissipative source to find:
\begin{equation}
\rho_R(t) = \frac{3m_p^2}{4t^2}\left[1+\frac{1}{t}\left(\frac{\sqrt{3}m_p}{2\rho_{R_0}^{1/2}}-t_0\right)\right]^{-2}~.
\end{equation}
At sufficiently large $t$ the radiation becomes insensitive to its initial value and shows an attractor behaviour tending towards $\rho_R(t) = 3m_p^2/4t^2$. This explains why the radiation approaches the standard evolution given by the dotted red line in Figs. \ref{domination} and \ref{no-domination} at late times after dissipation (particle production) becomes irrelevant. However, the increase in the Hubble parameter during the phase transition makes the universe expand by a larger factor than in the standard radiation domination scenario. In Figure \ref{naexpansion} we show the evolution of the Hubble parameter and radiation energy density relative to a standard radiation-dominated universe, for different numbers of fermion species. As one can easily conclude, increasing $N_F$ enhances the effect of dissipation and the relative change of $H$ and $\rho_R$.

\begin{figure}[htbp]
\includegraphics[scale=0.5]{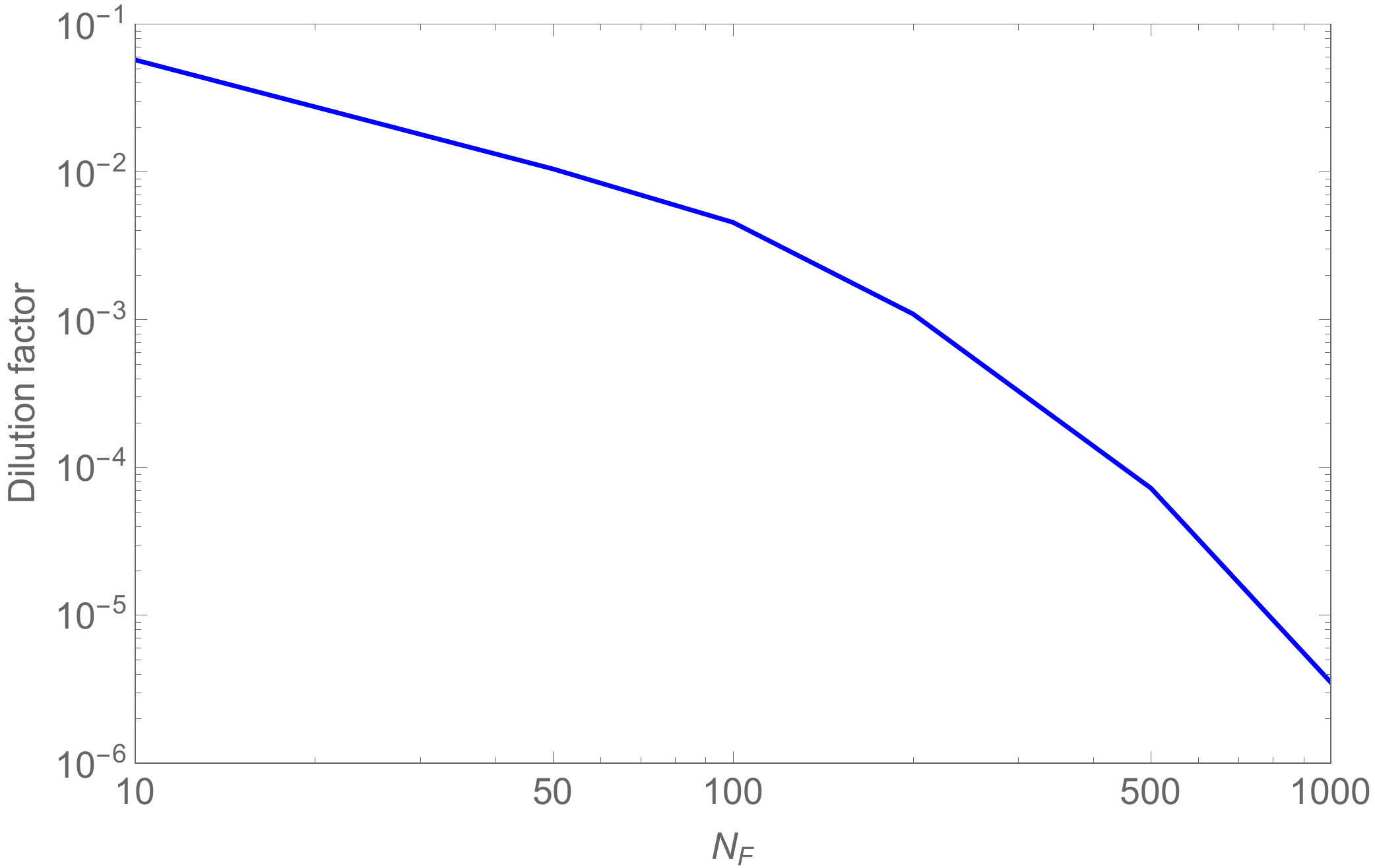}
\caption{Dilution factor for frozen relics during a phase transition as a function of the number of fermions coupled to the Higgs field, for $\lambda=10^{-2}$, $v=10^{15}$ GeV, $g^2 = 1/N_F$ and $h=0.1$.}
\label{dilution}
\end{figure}

The additional expansion will have a diluting effect on any decoupled particle species, for which the number density redshifts as $n\propto a^{-3}$. This includes e.g.~topological relics such as monopoles or thermal relics such as gravitinos generated prior to the phase transition. Considering an initial time $t_i$ before the critical temperature is reached and a final time $t_f$ after the field has settled into the symmetry breaking minimum, we have for a generic decoupled species:
\begin{equation} \label{number_density}
n_f= n_i\exp\left[-3\int_{t_i}^{t_f}Hdt'\right]~.
\end{equation}
Assuming no changes in the number of relativistic degrees of freedom, the entropy density of the heat bath before and after the phase transition is related by $s_f/s_i= (T_f/T_i)^3$. This implies that the number density-to-entropy ratio of the decoupled species is diluted by a factor:
\begin{equation} \label{dilution_factor}
{n_f/s_f\over n_i/s_i}= {\exp\left[-3\int_{t_i}^{t_f}Hdt'\right]\over (T_f/T_i)^3}~.
\end{equation}
In Fig. \ref{dilution} we show numerical results for this dilution factor as a function of the number of fermion species, showing that stronger dissipative effects  lead to a more significant dilution of dangerous relics, by enhancing either the maximum value of $\Omega_\phi$ attained or the duration of the late period of warm inflation. For example, observational constraints on the abundance of GUT monopoles require at least $n_M/s\lesssim 10^{-11}$ \cite{Kolb:1990vq,Preskill:1984gd}, so it is unlikely that a single phase transition subsequent to monopole formation can yield the required dilution factor unless a very large number of dissipative channels is involved. Even if a complete dilution cannot be achieved, this may, for example, alleviate the bounds on the reheating temperature after inflation concerning the overproduction of gravitinos. In particular, since $\Omega_{3/2} \propto T_R$ (see e.g.~\cite{Pradler:2006qh}), bounds on $T_R$ will increase by the inverse of the dilution factor in Eq.~(\ref{dilution_factor}). Furthermore, the cumulative effect of several different stages of symmetry breaking may potentially result in a significant dilution factor that should be taken into account.

In summary, we have shown that dissipative effects during a cosmological phase transition may have a significant effect on the cosmic history. By overdamping the motion of the associated Higgs field, dissipation not only prevents oscillations about the symmetry-breaking minimum but also leads to a period of slow-roll and potentially late-time warm inflation. The energy density in the field and the entropy produced by dissipative effects will also generically increase the amount of Hubble expansion during the phase transition and parametrically dilute the abundance of frozen relics. 

One or more short periods of late time warm inflation during phase transitions could have significant observational effects. On the one hand, their existence implies that the main period of inflation can be considerably shorter than the overall 50-60 e-folds of accelerated expansion required by the observed flatness and homogeneity of the universe. This will therefore change observational predictions for large scales, along the lines suggested in \cite{Kawasaki:2009hp} for the case of thermal inflation. On the other hand, small scale perturbations will be generated during these periods, although they should be well within the horizon today and hence potentially too damped to be studied in galaxy surveys or CMB observations. Although this requires further inspection and a detailed study that is outside the scope of this work, we nevertheless emphasize that dissipation will modify the evolution of fluctuations, therefore yielding distinct observational predictions from a period of thermal inflation. Since both thermal and dissipative (warm) inflation may occur within the same phase transition, it would be interesting to explore the combined effects  of these two types of inflationary expansion on the spectrum of cosmological perturbations.

%%%%%%%%%%%%%%%%%%%%%%%%%%%%%%%%%%%%%%%%%%%%%%%%%%%%%%%%%%%%%%%%%%%%%%%
%%%%%%%%%%%%%%%%%%%%%%%%%%%%%%%%%%%%%%%%%%%%%%%%%%%%%%%%%%%%%%%%%%%%%%%%%%%%%%%%%%%%%%%%%%%%%%%%%%%%%%%%%%%%%%%%%%%%%%%%%%%%%%%%%%%%%%%%%%%%%%%%%%%%%%%%%%%%%%%%%%%%%%%%%%%%%%%%%%%%%%%%%%%%%%%%%%%%%%%%%%%%%%%%%%%%%%%%%%%%%%%%%%%%%%%%%%%%%%%%%%%%%%%%%%%%%%%%%%%%%%%%%%%%%%%%%%%%%%%%%%%%

\section{Dissipative baryogenesis and leptogenesis}
\label{Leptogenesis}

As we have seen in the previous section, dissipation may have a significant effect in the dynamics of a cosmological scalar field in the process of spontaneous symmetry breaking. Significant effects arose in this case when the field became a non-negligible component of the energy balance in the universe, either itself increasing the Hubble rate or leading to a significant entropy production. In this section, we will consider an effect of dissipation that may occur even when the dissipating scalar field carries a very small fraction of the energy in the Universe and plays a subdominant role in entropy production.

Dissipation leads to the production of particles within the heat bath to which a dynamical scalar is coupled to, continuously disturbing its equilibrium. The degrees of freedom within the heat bath will {\it a priori} include the SM particles and their anti-particles, as well as potentially dark matter particles and other beyond the SM species. The rate at which each particle species is produced is related to its fractional contribution to the dissipation coefficient, as explicitly shown in \cite{Graham:2008vu}. It is then natural to envisage scenarios where particles and anti-particles are produced at different rates by a dissipating scalar field, necessarily involving interactions that violate baryon/lepton number as well as the C and CP symmetries, according to the conditions first established by Sakharov \cite{Sakharov:1967dj}. This was first explored in the context of warm inflation in a mechanism dubbed {\it warm baryogenesis} \cite{BasteroGil:2011cx}, where the same interactions responsible for damping the inflaton's motion and sustaining a radiation bath during inflation were shown to yield a significant baryon asymmetry, parametrically within the observed window. 

Here, we will show that {\it dissipative baryogenesis} is a much more general mechanism that may occur in the dynamics of any cosmological scalar field with non-equilibrium dissipative dynamics and interactions satisfying the Sakharov conditions. We will illustrate this by looking at a concrete example based on the interactions employed in standard thermal leptogenesis scenarios with right-handed neutrinos and which is naturally motivated within GUT models. Although dissipative baryo/leptogenesis will occur in several different dynamical regimes, we will focus on low-temperature dissipative models to explicitly show that the production of a lepton asymmetry does not require temperatures above the right-handed neutrino mass threshold as in the standard thermal scenarios. Our example further shows that no symmetries need to be restored in the early universe to generate the observed baryon asymmetry, thus avoiding the several potential cosmological problems that this may cause.

We will first consider the relevant particle physics interactions and describe how they lead to dissipative effects that may produce more particles than their anti-particles, and afterwards describe the dynamics of dissipative baryogenesis in the radiation-dominated era.

%%%%%%%%%%%%%%%%%%%%%%%%%%%%%%%%%%%%%%%%%%%%%%%%%%%%%%%%%%%%%%%%%%%%%%%%%%%%%%%%%%%%%%%%%%%%%%%%%%%%%%%%%%%%%%%%%%%%%%%%%%%%%%%%%%%%%%%%%%%%%%

\subsection{Interactions and dissipative particle production rates}

Leptogenesis is amongst the most popular models for the generation of a cosmic baryon asymmetry  \cite{Fukugita:1986hr, Kusenko:2014uta,Davidson:2008bu,Kusenko:2014lra,Luty:1992un,Cline:2006ts}. In the simplest models, it is based on the out-of-equilibrium decays of heavy right-handed Majorana neutrinos, which violate lepton number as well as C and CP. The resulting lepton asymmetry is later on converted into a baryon asymmetry by electroweak sphaleron processes \cite{'tHooft:1976up}, which conserve $B-L$ but not the two global charges independently. Heavy right-handed neutrino singlets are the simplest addition to the SM particle content, yielding light neutrino masses through the seesaw mechanism, thereby providing an interesting connection between cosmology and low-energy particle physics.

Right-handed neutrinos also fit nicely within the  $\mathbf{16}$ fundamental representation of the $SO(10)$ GUT gauge group and their large Majorana mass required by the seesaw mechanism can in this case be generated by the vev of a Higgs field in the $\mathbf{126}$ representation \cite{Babu:1992ia}. It is thus natural to consider the cosmological dynamics of this scalar field, which to our knowledge remains unexplored, including in particular the dissipative effects associated with its couplings to right-handed neutrinos. We will then consider a supersymmetric model where the relevant interactions are encoded in the superpotential:
\begin{equation}\label{lepw}
W = \frac{1}{2}g_a \Phi N^c_aN^c_a + y_{ai} N_a^cH_uL_i + f(\Phi)~,
\end{equation}
which involves the right-handed neutrino superfields, $N_a$, as well as the SM lepton and Higgs doublet superfields, $L_i$ and $H_u$, respectively. We consider three neutrino and lepton generations denoted by the indices $a$ and $i$ and note that $SU(2)$ gauge indices are implicit in the superpotential. The chiral superfield $\Phi$ can be identified with the scalar direction within the $\mathbf{126}$ representation of $SO(10)$ that gives a Majorana mass to the right-handed neutrinos, as discussed above, or more generally as a SM singlet with self-interactions encoded in the analytic function $f(\Phi)$. Without loss of generality, we will take its scalar vev as a real field $\langle \Phi \rangle = \phi/\sqrt{2}$.

Dissipation of the scalar field's energy will in this case proceed through the excitation of the right-handed neutrinos and their scalar superpartners in the cosmic heat bath and their subsequent decay into the MSSM (s)leptons and Higgs(inos). The cosmological evolution of the $\phi$ field will depend on its potential, given by $|f'(\phi)|^2$, and crucially on its behavior during inflation. As anticipated above, we will be mainly interested in studying the regime where right-handed neutrinos are too heavy to be thermally produced and therefore standard leptogenesis scenarios  are inefficient. This is natural in scenarios where the field is light and hence overdamped during inflation, either remaining frozen at some potentially large initial value or driven towards a large vev by de Sitter fluctuations. In the low-temperature regime where the reheat temperature after inflation is below the right-handed neutrino mass threshold, dissipation proceeds through the excitation of virtual modes in the heat bath as discussed earlier in this work. Scalar modes, in this case the right-handed sneutrinos decaying dominantly into sleptons and Higgs bosons, yield the leading contribution to the dissipation coefficient  \cite{BasteroGil:2012cm}, which is given approximately by: 
\begin{equation} \label{dissip_lepto}
\Upsilon =C_{\phi}\frac{T^3}{\phi^2}, \hspace{1cm} C_{\phi} \simeq  \frac{1}{16\pi}\sum_{a,i}y_{ai}y_{ai}^*~.
\end{equation}
This coefficient therefore determines the overall entropy production rate in the form of MSSM particles produced in the thermal bath by the decays of the virtual right-handed sneutrinos. These decays violate lepton number, since the Majorana mass term precludes a consistent assignment of $L$ to $N_a^c$, and may also violate C and CP if the Yukawa coupling matrix has non-trivial phases, which is possible for at least three matter generations. If this is the case then out-of-equilibrium dissipation will naturally induce an overabundance of sleptons over anti-sleptons in the heat bath (or vice-versa, although we will assume this to be the case). 

\begin{figure}[htbp]
\centering
\includegraphics[scale=0.5]{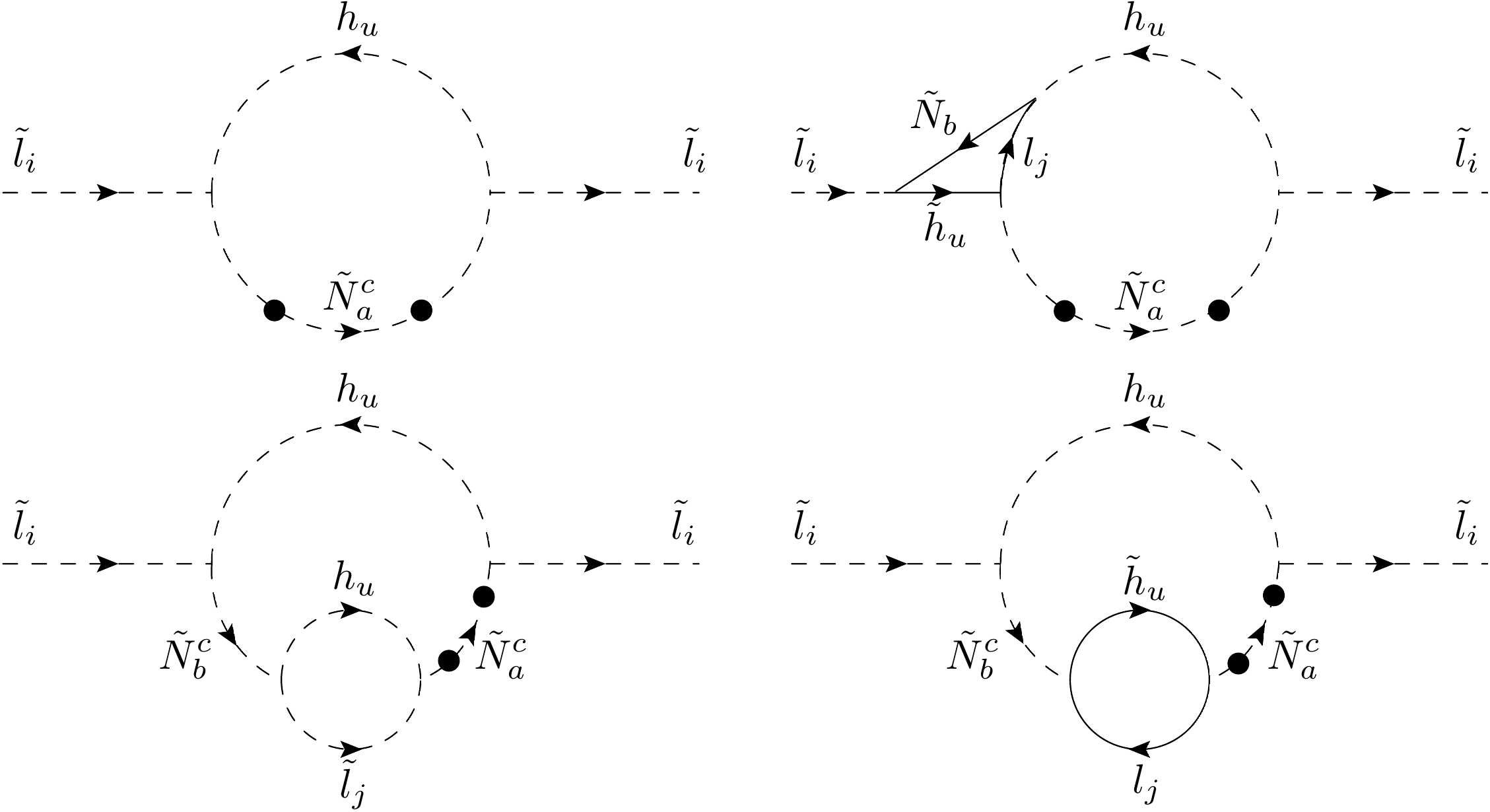}
\caption{The 1-loop and 2-loop diagrams contributing to the slepton self energy. Black circles indicate background field (i.e. right-handed sneutrino mass) insertions.}
\label{selfenergies}
\end{figure}

The rate at which sleptons and anti-leptons are produced can be computed from the imaginary part of their self-energies, following the generic procedure first described in \cite{Graham:2008vu}:
\begin{equation} \label{lepton_self_energy}
\dot{n}(p) = \text{Im}\left[2\int_{-\infty}^t dt_2\frac{e^{-i\omega_p(t-t_2)}}{2\omega_p}\Sigma_{21}(p,t,t_2)\right]~.
\end{equation}
Integrating over the 3-momentum $p$ and summing over the energies of all the light particle species yields a source term for the radiation energy density $\Upsilon \dot\phi^2$ corresponding to the dissipation coefficient in Eq.~(\ref{dissip_lepto}). We are, however, interested in the difference between the slepton and anti-slepton production rates, which as we will show is a sub-leading effect compared to the overall dissipative entropy production. To compute the slepton self-energies we first consider the relevant scalar and Yukawa interactions resulting from the superpotential in Eq. (\ref{lepw}), which are given by:
\begin{eqnarray} \label{lepto_int}
\mathcal{L}_s  &=& \frac{g_a^2}{2}\phi^2|\tilde{N}_a|^2 + \frac{g_a\phi}{\sqrt{2}}y_{aj}\tilde{N}_a^*h_u \tilde{l}_j + h.c \\
\mathcal{L}_{Y}  &=& \frac{1}{2}y_{ai} \left( \tilde{N}_a^c \bar{\tilde{h}}_u P_L l_i + \tilde{N}_a^c \bar{l}_iP_L \tilde{h} + h_u \bar{l}_iP_L N_a + h_u \bar{N}_aP_L l_i + \tilde{l}_i\bar{\tilde{h}}_uP_L N_a + \tilde{l}_i\bar{N}_aP_L \tilde{h}_u + h.c\right)~.
\end{eqnarray}
In Fig. \ref{selfenergies} we show the leading 1- and 2-loop diagrams contributing to the slepton self-energies. We note that, even though dissipation is dominated by scalar modes, at 2-loop order fermions and scalars will give comparable contributions to the slepton self energy. It is also interesting to note that even though the final asymmetry is independent of the lepton number assignment chosen for the right-handed neutrinos, this choice determines which diagrams actually exhibit $L$-violation. For example, the Yukawa sector always violates $L$, whereas scalar interactions preserve it for $L(N_a^c) = 1$. In scenarios where $B-L$ is a spontaneously broken gauge symmetry, the $N_a^c$ superfield will have lepton number $-1$ and $L$ is violated by both types of interactions.

Analogously to standard leptogenesis scenarios, CP violation arises only through the interference between the leading and next-to-leading diagrams. The leading diagram corresponds in this case to the top-left diagram in Fig. \ref{selfenergies}, the imaginary part of which yields the (tree-level) decay of the right-handed sneutrinos. The slepton self-energies can be computed using standard thermal field theory techniques and we refer the reader to \cite{Graham:2008vu, BasteroGil:2011cx} for more technical details. Slepton and anti-slepton self-energies are related by charge conjugation and we obtain for the difference between the self-energies, to leading order:
\begin{equation} \label{self-energy-diff}
\Delta\Sigma_{21}^{\tilde{l}_i} = \frac{3}{16\pi}\sum_{a,b}\sum_j \int dp\int dk C(p,k) \left(\frac{T}{m_b}\right)^4\frac{m_b}{m_a}\text{Im}(y_{bi}y_{aj}^*y_{bj}y_{ai}^*)~,
\end{equation} 
where $m_a=g_a\phi/\sqrt{2}$ are the right-handed sneutrino masses to leading order,  assuming the MSSM Higgs and sleptons have zero or at least negligible expectation values at this stage in the cosmological evolution. We note that once the sum over all heavy sneutrino and light field generations is performed, only the diagrams involving fermionic propagators contribute to the self-energy difference above. The factor $C(p,k)$ is common to the self-energy of all the different particle species and therefore determines the overall dissipation coefficient given above. Adding the self-energies of all particle species we obtain to leading order the following overall radiation production rate:
\begin{equation} \label{self-energy_total}
\Sigma^R_{21} = 8\sum_a\sum_j\int dp\int dk\, C(p,k)y_{aj}y_{aj}^*~.
\end{equation}
The relative rate at which a lepton asymmetry is produced by dissipation can then be obtained by taking the quotient of Eqs.~(\ref{self-energy-diff}) and (\ref{self-energy_total}), yielding:
\begin{equation} \label{lepton_rate}
r_L \equiv \frac{\dot{n}_L}{\dot{n}_R^d} \simeq \frac{3}{64\pi}\frac{1}{\sum_{a} (yy^{\dagger})_{aa}}\sum_{a\ne b} \left(\frac{T}{m_b}\right)^4\frac{m_b}{m_a}\mathrm{Im}\left[(yy^{\dagger})^2_{ba}\right]~,
\end{equation}
where the sum over light fields running in the loop is implicit. Note that, as in thermal leptogenesis, a non-vanishing asymmetry can only be produced if at least two of the right-handed sneutrinos are non-degenerate, thus requiring distinct  $g_a$ couplings in the superpotential (\ref{lepw}).

A couple of important properties of the asymmetry production rate should be emphasized. Firstly, the dissipation coefficient in Eq.~(\ref{dissip_lepto}) is independent of the $g_a$ couplings to leading order, so that all three right-handed sneutrino species will be virtually excited by the motion of the $\phi$ field and contribute to the lepton asymmetry. This is in contrast to thermal scenarios, where the out-of-equilibrium decay of the lightest right-handed (s)neutrino will give a dominant contribution. Secondly, the asymmetry production rate is suppressed by $(T/m_a)^4\ll 1$, which is associated with the fact that the right-handed sneutrinos are only virtually excited, as opposed to thermal leptogenesis scenarios. This means that while in the latter mechanism one must consider small couplings and CP violating phases to yield the observed baryon asymmetry, in dissipative leptogenesis a small baryon-to-entropy ratio can result solely from the low temperature suppression. We note that the leading scalar loop diagrams contributing to the asymmetry are only suppressed by a factor $(T/m_a)^2$, but as mentioned above their overall contribution cancels out when summing over the different generations. This is a specific feature of the interaction structure considered in leptogenesis, with a single type of decay channel for the heavy right-handed sneutrinos, so that in more general models of dissipative baryogenesis, such as the one considered in \cite{BasteroGil:2011cx}, the asymmetry production rate will be larger.

The light neutrino mass hierarchy inferred from experimental bounds motivates considering a hierarchical structure in the right-handed neutrino sector as well, e.g.~$g_1\ll g_2 \ll g_3$. In this case the asymmetry production rate reduces to:
\begin{equation} \label{lepton_rate_light}
r_L \simeq \frac{3}{64\pi}\left(\frac{T}{m_1}\right)^4 \frac{1}{\sum_a(yy^{\dagger})_{aa}}\sum_{a\ne 1}^{N_R}\left(\frac{m_1}{m_a}\right) Im[(yy^{\dagger})_{1a}^2]~,
\end{equation}
which is suppressed relative to its thermal leptogenesis counterpart by a factor $(1/8) (T/m_1)^4$, as well as the fact that in the latter case only the lightest right-handed neutrino contributes to the factor $\sum_a(yy^{\dagger})_{aa}$ in the denominator which corresponds to the overall entropy production rate. To simplify our dynamical analysis of dissipative leptogenesis, we collect all couplings and mass differences into an effective parameter $\epsilon$, such that:
\begin{equation}
r_L \simeq \epsilon y^2  \left(\frac{T}{m_1}\right)^4~,
\end{equation}
where we assumed that the Yukawa couplings have roughly the same magnitude $y$.

%%%%%%%%%%%%%%%%%%%%%%%%%%%%%%%%%%%%%%%%%%%%%%%%%%%%%%%%%

\subsection{Dynamics of the lepton asymmetry generation}

Having determined the rate at which lepton number is produced by dissipation, we will now consider the dynamics of the scalar field $\phi$, which is coupled to the evolution of the overall entropy and lepton number density via the system of equations:
\begin{gather} \label{lepto_eqs}
\ddot{\phi}+(3H+\Upsilon)\dot{\phi} + V'(\phi)=0~, \\
\dot{s}+3Hs = \frac{\Upsilon\dot{\phi}^2}{T}~, \hspace{2cm}
\dot{n}_L + 3Hn_L = \frac{45\zeta(3)}{2\pi^4}\frac{g_L}{g_*}\frac{\Upsilon\dot{\phi^2}}{T} r_L~,
\end{gather}
where $g_L$ is the number of relativistic degrees of freedom with non-vanishing lepton number, for which we will take the MSSM value $g_L = 33.75$ as a reference, as well as the associated $g_*=228.75$ for the overall number of relativistic species. We will consider the evolution in the low-temperature dissipative regime, where the dissipation coefficient takes the form in Eq.~(\ref{dissip_lepto}). We will assume that the Yukawa couplings have roughly the same magnitude for all three generations, such that $C_{\phi}\simeq 9y^2/16\pi$, although this assumption is not crucial for our subsequent analysis. The lepton number density is sourced, as computed in the previous section, by a fraction $r_L\Upsilon$ of the overall dissipation coefficient.

We are interested in the evolution of the field $\phi$ in a regime where it has a large vev, such that right-handed (s)neutrinos have a large mass and cannot be produced on-shell. This implies that, as opposed to the example considered in the previous section, we assume that there is no symmetry restoration after inflation. As a concrete example, we take the simple symmetry breaking potential of the previous section, given in Eq.~(\ref{Higgs_potential}), although thermal mass corrections will always be Boltzmann-suppressed in the regime that we are interested in exploring. 

 We note that if the field $\phi$ comes to dominate over the radiation energy density, or at least attains a significant relative abundance, then a sizeable lepton asymmetry can be produced, which is the case of the warm baryogenesis mechanism during inflation \cite{BasteroGil:2011cx}.
We will show, however, that the observed baryon asymmetry can also be produced when the field is sub-dominant and dissipation does not contribute significantly to the overall entropy of the universe.

It is convenient to express the lepton number density in terms of the lepton-to-entropy ratio, $Y_L \equiv n_L/s$, which becomes constant once a lepton asymmetry stops being efficiently produced. We then have: 
\begin{equation} \label{yield}
\dot{Y}_L = C_L \frac{T^3}{\phi^6}\dot{\phi}^2, \hspace{1cm} C_L = \frac{90\zeta(3)}{\pi^4} \frac{g_L}{g_*} \frac{C_{\phi}}{C_s}\epsilon {y^2\over g^4}~,
\end{equation}
where  $C_s = 2\pi^2 g_* /45$. The evolution of $Y_L$ will then be determined by the dynamics of the $\phi$ field. We assume the field's self-coupling $\lambda$ is sufficiently small for it to be overdamped during inflation, $m_\phi^2=\lambda^2(3\phi^2-v^2)\ll H_{inf}^2$, where $H_{inf} \lesssim 10^{14}\ \mathrm{GeV} $ from the recent CMB upper bounds on the tensor-to-scalar ratio obtained by the Planck satellite \cite{Ade:2013uln}. De Sitter fluctuations will then lead to a distribution of field values $\phi_i\lesssim M_P$ in different patches of the inflationary universe at the start of the radiation era, and which will typically be displaced from the minimum of the potential at $|\phi|=v$.

As it evolves towards the minimum, the field will feel the effects of both Hubble damping and dissipative friction. The latter will play a significant role for:
\begin{equation} \label{Q_baryo}
Q= {\Upsilon\over 3H}\simeq \sqrt{10\over \pi^2}{C_\phi\over \sqrt{g_*}} {T M_P\over \phi^2}\gtrsim 1~,   
\end{equation}
where we have used the standard relation between the Hubble rate and the ambient temperature in a radiation-dominated universe. On the one hand, if the field rolls towards the minimum from $|\phi_i|<v$, $Q$ will necessarily decrease in time and so dissipation can at most have a significant effect during the earlier stages of the evolution. On the other hand, for $|\phi_i|>v$, $Q$ may increase as the field value decreases, in particular if it overshoots the minimum and attains a small value during the first oscillation. In any case dissipation can only have a transient effect, since asymptotically the field will settle at the minimum and $Q$ will decrease with the temperature.

For simplicity, we will focus on scenarios where dissipation plays no significant role in the field dynamics. This is, in particular, the case for a large field vev $v\lesssim M_P$ and initial displacements $\Delta\phi_i=|\phi_i-v|\lesssim v$, for which $Q\lesssim T/M_P\ll 1$. In the standard seesaw mechanism the right-handed (s)neutrino masses are related to the light neutrino masses via:
\begin{equation} \label{field_vev}
m_{\tilde{N}}={gv\over \sqrt{2}} \simeq  10^{15}y^2\left({0.1\ \mathrm{eV}\over m_\nu}\right) \ \mathrm{GeV} ~, 
\end{equation}
so that $v\lesssim M_P$ implies $g\gtrsim 10^{-3}y^2 (0.1\ \mathrm{eV}/ m_\nu)$. Under these conditions the field oscillations are well described by:
\begin{equation} \label{field_sol}
\phi\simeq  v +\Delta\phi_i\left({t\over t_i}\right)^{-3/4}\cos\left(m_\phi(t-t_i)\right)~,
\end{equation}
where $m_\phi=\sqrt{2}\lambda v$ is the field mass at the minimum, $\Delta\phi_i\equiv \phi_i-v$ and $t_i= (2H_i)^{-1}\sim (2m_\phi)^{-1}$ is the time at which the field becomes underdamped and effectively starts oscillating. We may then substitute this into Eq.~(\ref{yield}) to estimate the lepton-to-entropy ratio produced by dissipation as the field oscillates. We note that since the field velocity is small before the onset of oscillations, no significant lepton number will be produced until the field becomes underdamped. Taking the average field value $\langle \phi\rangle= v$ and the average field velocity $\langle \dot\phi^2\rangle\simeq m_\phi^2 \Delta\phi_i^2(t/t_i)^{-3/2}/2$, we can then integrate Eq.~(\ref{yield}) from $t=t_i$ to obtain asymptotically:
\begin{equation} \label{final_yield}
Y_L \simeq {C_L\over 4 g_*^{3/4}} \left({45\over 2\pi^2}\right)^{3/4}\left(M_P t_i\right)^{3/2} {m_\phi^2\Delta\phi_i^2 \over v^6}{1\over t_i^2}~.
\end{equation}
For $g_*=228.75$ and $g_L=33.75$ in the MSSM, and taking into account the relation between the asymptotic baryon and lepton numbers after conversion by sphaleron processes, $B_f = - (8/23) L_i$ \cite{Cline:2006ts}, this yields:
\begin{equation} \label{final_asym_1}
\eta_s \simeq 0.1\left({m_\nu\over 0.1\ \mathrm{eV}}\right)^{2}\left({m_{\tilde{N}}\over 10^{15}\ \mathrm{GeV}}\right)^{1/2}\left({\Delta\phi_i\over v}\right)^2\left({m_\phi\over m_{\tilde{N}}}\right)^{5/2}\epsilon~.
\end{equation}
The baryon asymmetry thus depends parametrically on the ratio of the field and sneutrino masses, $m_\phi/m_{\tilde{N}}=2\lambda/g$. Adiabaticity of the dissipative process requires the field to move slowly compared to the sneutrino decay rate, $\dot\phi/\phi \sim m_\phi \ll \Gamma_{\tilde{N}}\simeq y^2 m_{\tilde{N}}/16\pi$. Additionally, for the asymmetry to be produced below the sneutrino mass threshold, we require $T< m_{\tilde{N}}$ at the onset of field oscillations, $H\sim m_\phi$. This then implies $m_\phi/m_{\tilde{N}}\lesssim m_{\tilde{N}}/M_P$, which is typically a stronger constraint than adiabaticity of the dissipative processes. Saturating this bound, we obtain for the final baryon asymmetry:
\begin{equation} \label{final_asym_2}
\eta_s \simeq 10^{-10}\left({m_\nu\over 0.1\ \mathrm{eV}}\right)^{2}\left({m_{\tilde{N}}\over 2\times 10^{15}\ \mathrm{GeV}}\right)^{3}\left({\Delta\phi_i\over v}\right)^2\left({\epsilon\over 0.05}\right)~.
\end{equation}
We thus see that the observed baryon asymmetry, $\eta_s\sim 10^{-10}$ \cite{Fields:2006ga}, can be obtained through adiabatic dissipation for sneutrino masses close to the GUT scale, which generate neutrino masses in the range suggested by atmospheric neutrino oscillations for $\mathcal{O}(1)$ Yukawa couplings \cite{Agashe:2014kda}. As opposed to standard leptogenesis models, the amount of CP-violation, parametrized by $\epsilon$, need not be very small in this case since the produced baryon asymmetry is naturally small. We note that Eq.~(\ref{final_asym_2}) is an estimate that is accurate up to $\mathcal{O}(1)$ factors, since the onset of field oscillations does not occur exactly for $H=m_\phi$, so the above values for the masses should be taken only as reference values. 

\begin{figure}[h!]
\centering
\subfigure{
\includegraphics[scale=0.4]{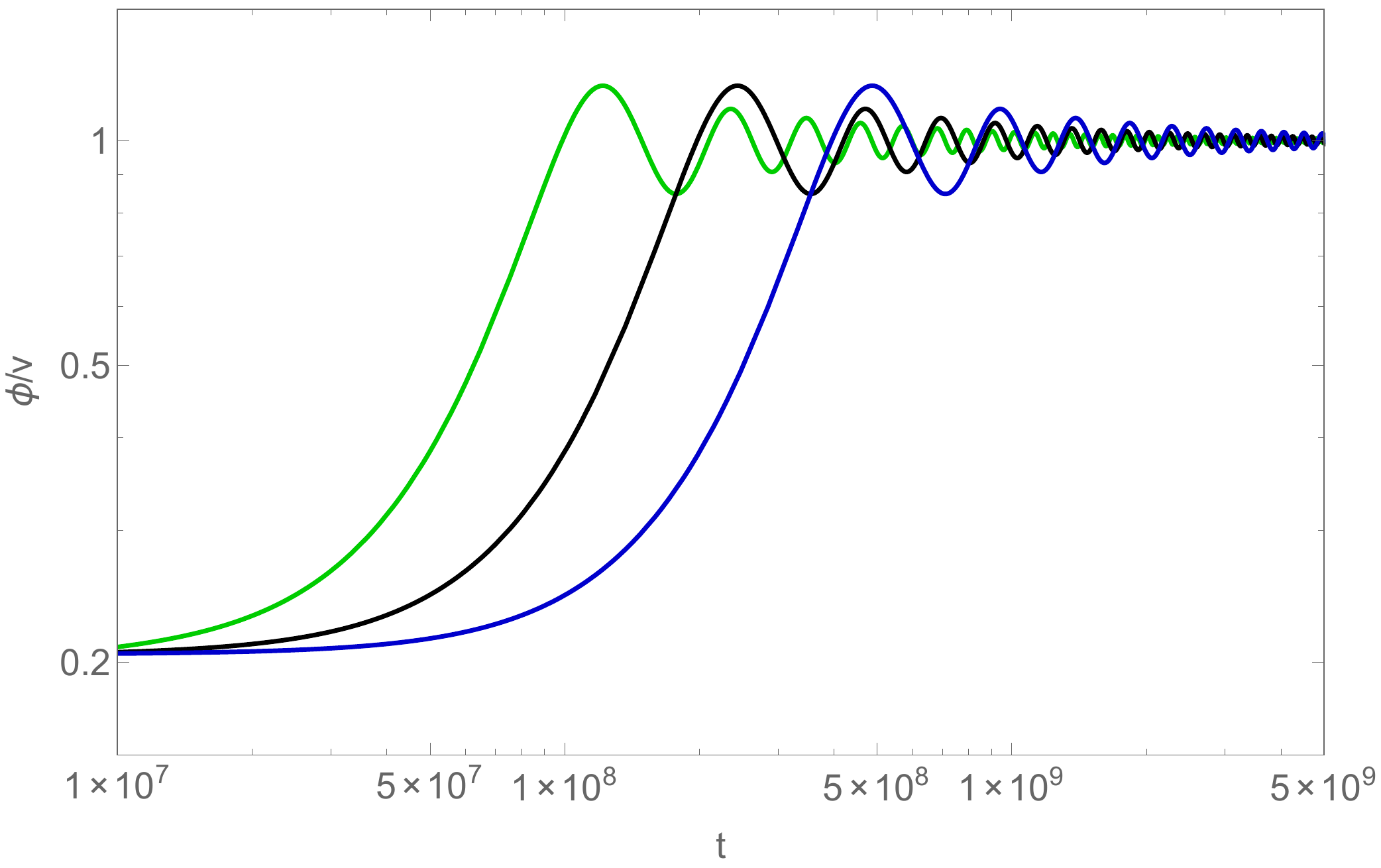} 
}
\subfigure{
\includegraphics[scale=0.4]{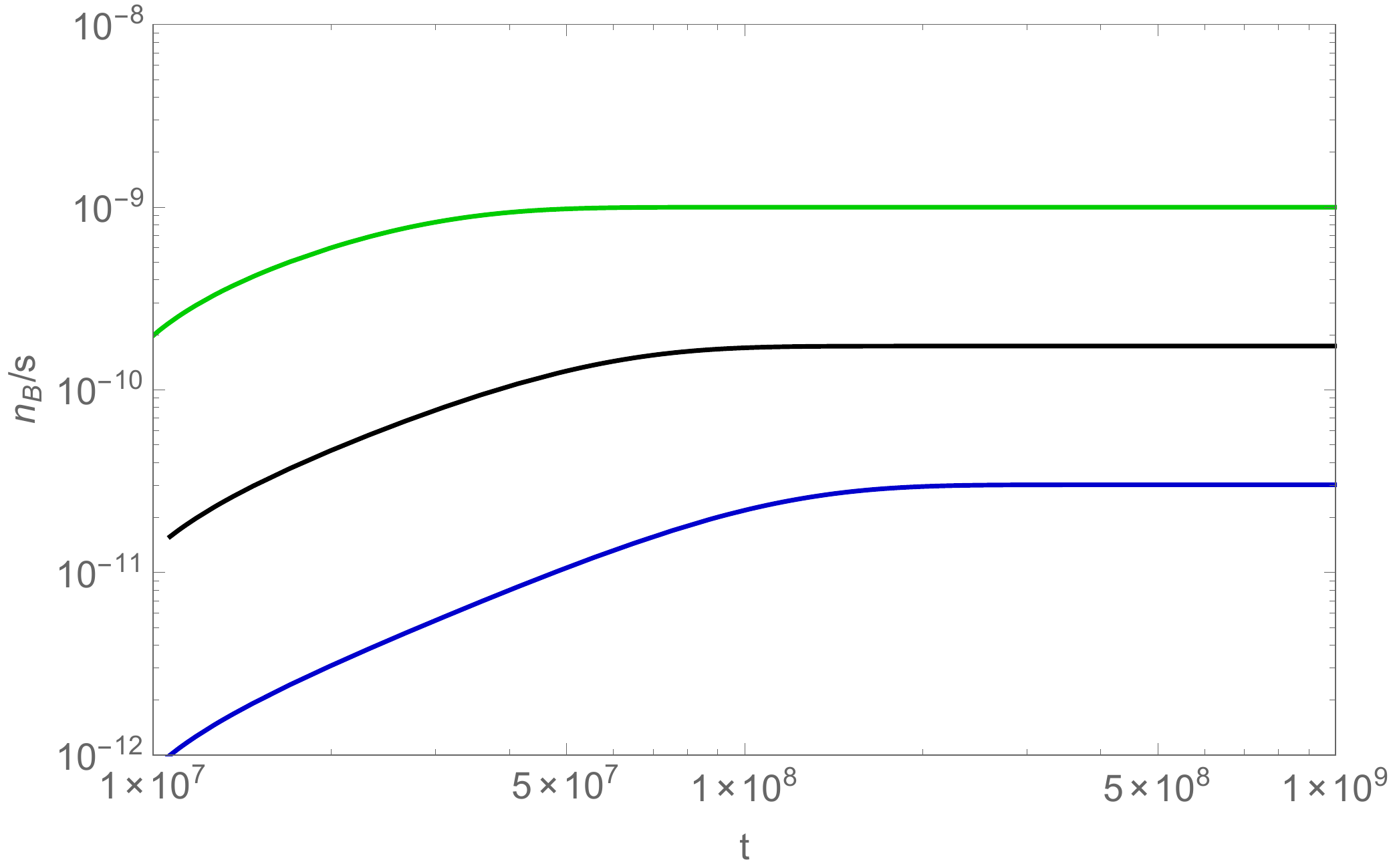}
}
\caption{Numerical results for the evolution of the field $\phi$ and the baryon asymmetry $\eta_s=n_B/s$, starting from a field value below the minimum at $v=10^{18}$ GeV. The quartic self-coupling $\lambda = 1 \times 10^{-8}, 2 \times 10^{-8}, 4 \times 10^{-8}$ for the blue, black and green curves, respectively. We have taken $g=10^{-3}$, $y=3$ and $\epsilon=1/64\pi$ in all cases. Time is given in Planck units.}
\label{smallphi}
\end{figure}

\begin{figure}[h!]
\centering
\subfigure{
\includegraphics[scale=0.4]{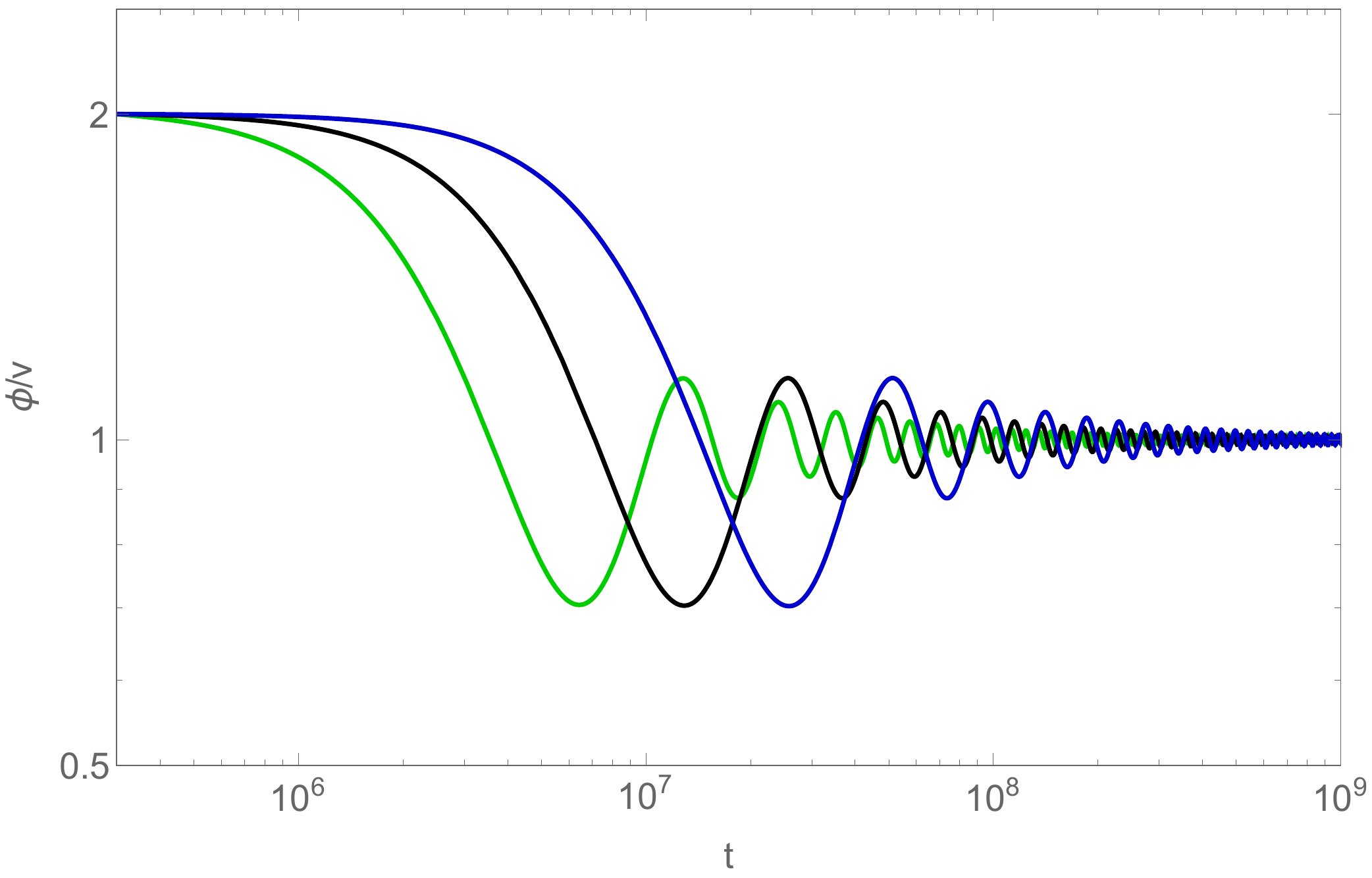} 
}
\subfigure{
\includegraphics[scale=0.4]{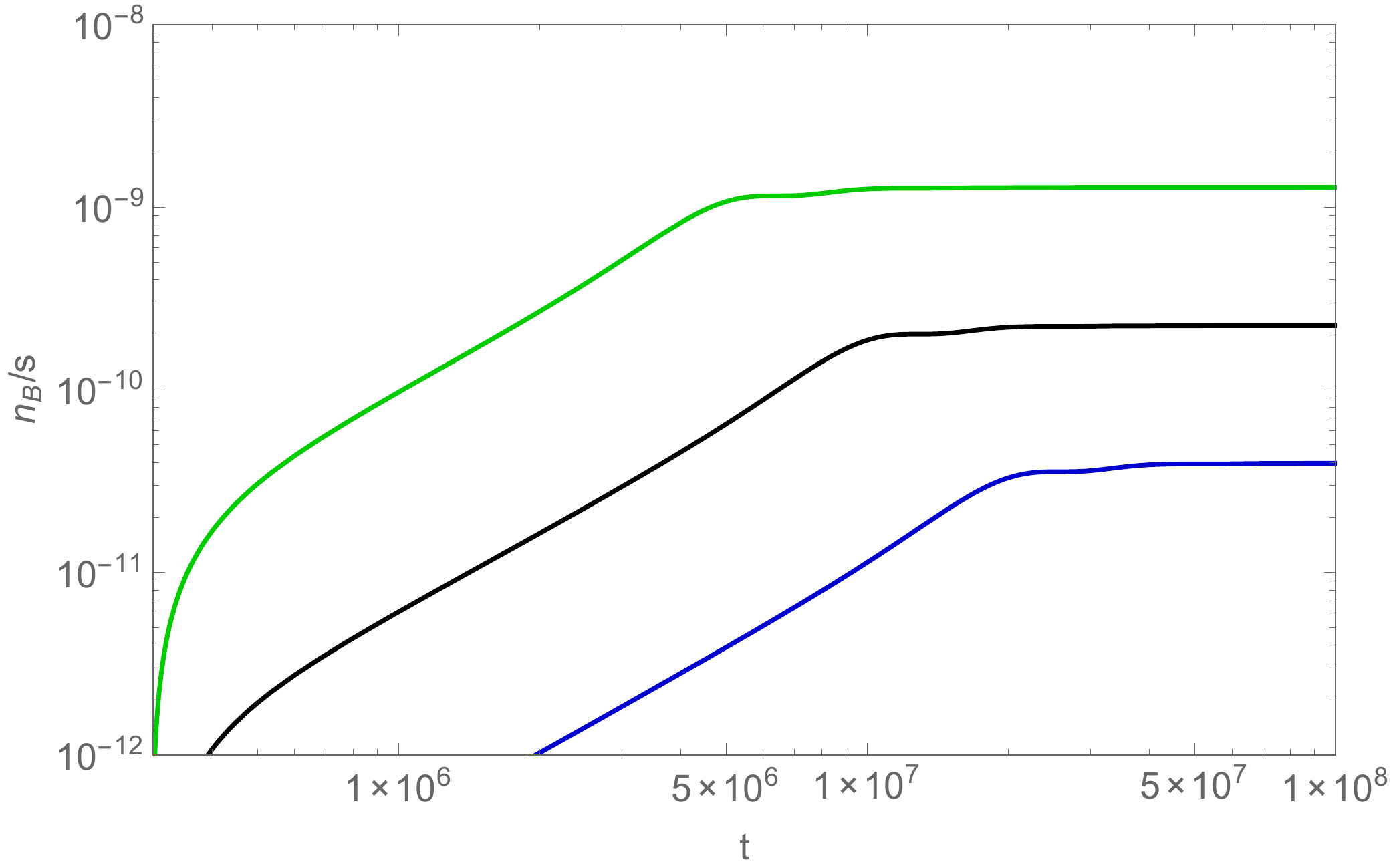}
}
\caption{Numerical results for the evolution of the field $\phi$ and the baryon asymmetry $\eta_s=n_B/s$, starting from a field value above the minimum at $v=M_P$. The quartic self-coupling $\lambda = 1 \times 10^{-7}, 2 \times 10^{-7}, 4 \times 10^{-7}$ for the blue, black and green curves, respectively. We have taken $g=10^{-3}$, $y=3$ and $\epsilon=1/64\pi$ in all cases. Time is given in Planck units.}
\label{largephi}
\end{figure}

The exact value of the produced asymmetry can be computed numerically, and in Figures \ref{smallphi} and \ref{largephi} we give examples for the numerical evolution of the field $\phi$ and the asymmetry $n_B/s$ in the regime considered above, for different values of the field mass parametrized by the self-coupling $\lambda$. In all examples shown, dissipation has a sub-dominant effect on the field evolution, as discussed above, and its main effect is the production of a baryon asymmetry. We have checked in all cases that the adiabatic condition is satisfied and that the temperature is below the sneutrino mass threshold at the onset of field oscillations.

We thus conclude that the produced asymmetry can have a range of values both below and above the observational window. Most of the lepton number is produced in the first few oscillations of the field, where the field velocity, and hence the dissipative lepton source, is larger, with the lepton-to-entropy ratio stabilizing within a few oscillation periods.

We note that, even though the adiabatic dissipation coefficient decreases with the temperature, and hence becomes negligible at late times, the full dissipation coefficient includes a zero-temperature part that corresponds to the standard decay width for an oscillating field \cite{Graham:2008vu}. This corresponds in the present scenario to the 4-body decay of the $\phi$ field into Higgs and slepton pairs mediated by virtual right-handed sneutrinos, since the latter's on-shell production is kinematically forbidden. As shown in \cite{Graham:2008vu}, this contribution is suppressed by $(m_\phi/T)^3$, as well as numerical factors, with respect to the adiabatic component.  We may thus safely neglect this contribution  in computing the lepton asymmetry, which is produced when $m_\phi\sim H \ll T$, bearing nevertheless in mind that this will lead to the decay of the $\phi$ field after it becomes non-relativistic at late times.

In the particular model of leptogenesis that we have considered, a lepton (and hence baryon) asymmetry is produced by the dynamical evolution of a scalar SM singlet that determines the Majorana mass of right-handed (s)neutrinos. This is, however, a much more general result and dissipative baryogenesis should occur in any scenario where fields whose decay violates the B/L- and C, CP-symmetries are coupled to (and acquire mass from) a dynamical scalar field, including e.g.~the $SU(5)$ model considered in Section II. Depending on the field masses and couplings, the observed baryon asymmetry may be entirely produced by off-shell dissipative effects, with no need for temperatures above the B-violating field mass threshold. This may then avoid symmetry restoration in the early universe and the production of dangerous thermal relics during the associated phase transitions. In addition, dissipative baryogenesis generically yields potentially observable signatures, as we describe below.

%%%%%%%%%%%%%%%%%%%%%%%%%%%%%%%%%%%%%%%%%%%%%%%%%%%%%%%%%%%%%%%%%%%%%%%%%%%%%%%%%%%%%%%%%%%%%%%%%%%%%%%%%%%%%%%%%%%%%%%%%%%%%%%%%%%%%%%%

\subsection{Isocurvature perturbations}

As obtained above, the baryon asymmetry that results from dissipative processes will depend on the initial field displacement from the true minimum of its potential. If, as we assumed earlier, the field is light during inflation, we then expect super-horizon quantum fluctuations $\langle\delta\phi_i^2\rangle= (H_{inf}/2\pi)^2$ in the initial field value. These will then result in fluctuations in the final baryon-to-entropy ratio and hence baryon isocurvature modes that can be tested with CMB observations. 

This is also a feature of the warm baryogenesis scenario during (warm) inflation \cite{BasteroGil:2011cx}, where both inflaton and temperature fluctuations generate baryon isocurvature modes. The main difference to the case analyzed in this work resides, firstly, in the fact that the field responsible for producing the baryon asymmetry never dominates the energy balance in the universe. Consequently, the resulting baryon isocurvature modes will be uncorrelated with the main (adiabatic) curvature perturbations sourced by the inflaton. Secondly, the baryon asymmetry is produced at the onset of field oscillations rather than in a slow-roll regime. Since this occurs when $H\sim m_\phi$, fluctuations in the ambient temperature will only delay or expedite the production of the baryon asymmetry, but they do not change its final value. From Eq.~(\ref{final_asym_2}), we have that:
\begin{equation} \label{iso_S}
S_B = \frac{\delta\eta_s}{\eta_s} = 2 \frac{\delta \phi_i}{\Delta\phi_i}~,
\end{equation}
such that we can write the relative contribution of uncorrelated baryon isocurvature modes to the CMB spectrum as:
\begin{equation} \label{iso_B}
B_B^2 = \frac{S_B^2}{P^2_{\zeta}} \simeq {r\over 2} \left(\frac{M_p}{\Delta\phi_i}\right)^2~,
\end{equation}
where $P_\zeta^2\simeq 2\times 10^{-9}$ is the amplitude of the adiabatic curvature perturbation spectrum \cite{Ade:2013uln} and $r$ is the tensor-to-scalar ratio. This gives a contribution to the total matter isocurvature power spectrum that is suppressed by the relative abundance of baryons $(\Omega_b/\Omega_m)^2B_B^2\simeq 0.03 B_B^2$. From the constraints posed by the Planck satellite on uncorrelated CDM isocurvature modes with a scale invariant spectrum at a comoving wavenumber $k_{low} =0.002\ \mathrm{Mpc}^{-1}$ \cite{Ade:2013uln}, we deduce the bound $B_B\lesssim 1.03$. The result above satisfies this bound for initial field displacements:
\begin{equation} \label{iso_bound}
\Delta\phi_i> 0.68 \sqrt{r} M_P~.
\end{equation}
If the tensor-to-scalar ratio is close to the current upper bound $r\lesssim 0.1$ \cite{Ade:2013uln}, this requires field displacements $\Delta\phi_i\gtrsim 0.2 M_P$. Although the value of the baryon asymmetry does not depend directly on the actual value of the field displacement, but rather on the ratio $\Delta\phi_i/v$, we have seen above that the observed baryon asymmetry can be entirely produced by off-shell dissipative effects for $\Delta\phi_i \sim v\sim M_P$. The bound above is thus consistent with the generation of the observed baryon asymmetry. On the other hand, low-scale inflationary models with $r\ll 0.1$ are consistent with initial field displacements parametrically below the Planck scale.

We note, however, that in supergravity models scalar fields may acquire masses parametrically close to the Hubble scale during inflation, and hence be driven to a local minimum that does not necessarily coincide with the low-energy global minimum \cite{Dine:1995kz}. In this case dissipation may also produce a baryon/lepton asymmetry as the field rolls towards the true minimum after inflation, although field masses of the order of the Hubble scale may somewhat change the dynamics. In these scenarios there will be, however, no significant field fluctuations on super-horizon scales, which makes them less appealing from the observational point of view.

Evidence for baryon isocurvature modes will nevertheless constitute a strong hint for dissipative baryogenesis, which is thus a testable mechanism. In addition, the particular case of dissipative leptogenesis considered above can be related to low-energy neutrino phenomenology, thus yielding two independent potential ways of probing the production of a baryon asymmetry.

%%%%%%%%%%%%%%%%%%%%%%%%%%%%%%%%%%%%%%%%%%%%%%%%%%%%%%%%%%%%%%%%%%%%%%%%%%%%%%%%%%%%%%%%%%%%%%%%%%%%%%%%%%%%%%%%%%%%%%%%%%%%%%%%%%%%%%%%
%%%%%%%%%%%%%%%%%%%%%%%%%%%%%%%%%%%%%%%%%%%%%%%%%%%%%%%%%%%%%%%%%%%%%%%%%%%%%%%%%%%%%%%%%%%%%%%%%%%%%%%%%%%%%%%%%%%%%%%%%%%%%%%%%%%%%%%%

\section{Conclusion}
\label{Conclusion}

Scalar fields are ubiquitous in the best-motivated extensions of the Standard Model of particle physics and their dynamics has in most cases a very significant cosmological impact. Since they generically interact with other matter and gauge degrees of freedom, dissipative effects are a crucial feature determining how scalar fields evolve in the cosmological heat bath. This leads to additional friction, entropy production and scalar field fluctuations. These effects are, in the leading adiabatic approximation, fully encoded in a single dissipation coefficient, which can be computed from the fundamental Lagrangian defining the properties and interactions of a given scalar field.

The study of dissipative effects has so far been mostly restricted to the early period of inflation, where dissipation may, in fact, completely change the inflaton dynamics and the associated generation of primordial curvature perturbations \cite{Bartrum:2013fia,Bartrum:2013oka,Bartrum:2012tg,Bastero-Gil:2014oga,Berera:1995ie,Berera:1996nv,Berera:1999ws,Hall:2003zp,Moss:2008yb,BasteroGil:2009ec,Berera:2002sp, Berera:2008ar}. More recently, a few studies have also begun to explore the importance of dissipation in the dynamics of reheating after inflation \cite{Mukaida:2012qn,Mukaida:2012bz,Drewes:2013iaa,Harigaya:2013vwa,BasteroGil:2010pb}, assuming it proceeds from a supercooled stage where dissipative dynamics plays a negligible role, and also in the dynamics of a curvaton field \cite{Mukaida:2014yia}.

It was the purpose of this work to set the stage for a much broader exploration of dissipative dynamics in the evolution of cosmological scalar fields, which in many cases only begins in the radiation-dominated era once the Hubble expansion rate and the ambient temperature have decreased sufficiently. A scalar field will typically find itself displaced from the absolute minimum of its effective potential after inflation and, in evolving towards it, the field will necessarily dissipate part of its energy into the ambient heat bath. Dissipation is thus, in particular, an inherent part of the process of spontaneous symmetry breaking and can modify the dynamics of the several phase transitions that may have occurred in the early cosmic history. The evaporation of a Bose condensate may also lead to an effective friction term along the lines suggested in \cite{Dymnikova:2001ga}.

A natural starting point for our study was to compute, for the first time, the dissipation coefficient inherent to different scalar fields in particle physics models of the early universe, and which we hope will be useful for future studies. We have, in particular, considered the electroweak Higgs field(s) in the SM and its minimal supersymmetric extension, the scalar singlet yielding the $\mu$-term in the NMSSM, the adjoint Higgs direction breaking the GUT $SU(5)$ group to the SM, and a SM singlet giving a Majorana mass to right-handed neutrinos (embedded e.g.~in $SO(10)$). For a given dynamical scalar, the dissipation coefficient takes different forms depending on the properties of the fields it is coupled to, namely their mass, spin, multiplicity and coupling constants. Dissipation proceeds generically through the excitation and decay of these fields and the associated coefficient takes different forms depending on whether on-shell or off-shell excitation is dominant. Note, for example,  that while on-shell modes resonantly enhance dissipation, their occupation numbers become Boltzmann-suppressed at temperatures below their mass threshold and virtual modes then yield the dominant contribution. Generically, dissipation coefficients depend both on the field value and the ambient temperature, thus constituting dynamical quantities.

We then proceeded to explore the dynamical impact of dissipative effects by solving the Langevin-like equation that determines the evolution of a cosmological scalar field, Eq.~(\ref{langevin}), in different scenarios. In this work we have focused mainly on the effects of dissipation and entropy production, although we have also briefly discussed the importance of the dissipative noise term, which we plan to explore in more detail in a future work. This term will be crucial, for example, in the initial stages of a cosmic phase transition, randomly kicking the associated Higgs field away from the unstable symmetric point. Fluctuation-dissipation will then play an important role in the formation of topological defects and also potentially in sourcing cosmic magnetic fields. There are various works in the literature which make statistical arguments for the distribution of seed magnetic domains. In a related context, statistical arguments are made on the initial distribution of a scalar field, such as in examining the initial condition problem of inflation \cite{Ijjas:2013vea,Mazenko:1985pu,Albrecht:1985yf,*Albrecht:1986pi}. The Langevin-like Eq.~(\ref{langevin}) provides a dynamical equation from which such distributions can be calculated rather than just argued statistically. 

The friction effects associated with dissipation will slow down a field's evolution towards the minimum of its potential and damp the amplitude of its oscillations about the minimum. This will prolong a cosmic phase transition, potentially even completely overdamping the associated Higgs field. One interesting outcome of our analysis is the possibility of dissipation keeping the field in a slow-roll regime close to the symmetric value, such that it drives a late period of warm inflation. This may last for a few e-folds, which may be sufficient to dilute dangerous thermal relics such as gravitinos or GUT monopoles formed at earlier stages. In fact, a period of thermal inflation above the critical temperature and a period of (dissipative) warm inflation below the critical temperature can occur within the same phase transition. Although they have a similar dilution effect, these two periods have inherently different field dynamics, as well as fluctuations, and we hope in the future to investigate more closely their potentially distinct observational impact in the CMB and/or matter power spectrum.

We have also observed that a slow-roll period typically ends with a Higgs field falling fast towards the symmetry-breaking minimum, in the process producing a significant amount of entropy. We have thus concluded that, even if the field does not become dominant during the slow-roll phase, a parametric dilution of unwanted relics will still occur as a combined result of this entropy production and enhancement of the expansion rate during the phase transition.

In this work, we have also shown that, even if dissipation does not significantly enhance friction or entropy production, it may nevertheless lead to a small but nevertheless crucial effect - the generation of a cosmological baryon asymmetry. This occurs due to the out-of-equilibrium nature of dissipative processes when B-/L- and C-/CP-violating interactions are involved. We have explicitly shown that the observed baryon asymmetry can be generated in a scenario of dissipative leptogenesis, where a SM scalar singlet giving a Majorana mass to right-handed (s)neutrinos excites the latter while rolling towards the minimum of its potential. This can occur at temperatures below the right-handed (s)neutrino mass threshold, with dissipation  dominantly exciting virtual modes. We expect this to be a generic feature, also found earlier in the context of warm baryogenesis during inflation \cite{BasteroGil:2011cx}, such that it is possible to produce the observed baryon asymmetry while avoiding symmetry restoration and subsequent production of topological defects. Furthermore, the baryon-to-entropy ratio generated through dissipation is generically field and temperature dependent, thus leading to baryon isocurvature perturbations that may be probed in the near future with CMB observations. If the observed asymmetry is produced by dissipation, we may thus hope to be able to test the violation of fundamental symmetries at high energies.

Our results show that going beyond the leading approximation of non-interacting adiabatic fluids can have a very significant impact in cosmology. Fluctuation-dissipation dynamics is present in any dynamical interacting system and, in particular, we have explicitly shown that this is the case for the SM Higgs field and for many other dynamical scalars present in its extensions. Significant dissipative effects are already a feature of near-equilibrium and near-adiabatic dynamics and thus should motivate further exploration of this topic, not only within this regime but also for more general non-equilibrium cosmological systems.

Up to now cosmology has been very successful in explaining almost all observations through the simple model of a thermalized universe that is expanding.  Phase transitions have been added to this picture to reconcile it with unified models of particle physics. There is a need to press beyond this simple picture and look with greater detail at the dynamics in the early Universe.  The early Universe is a many-body system with limited initial condition information. A theoretical treatment of it requires a statistical dynamical approach, which can extend on the thermal equilibrium hot big bang model. Phase transitions and other regimes of scalar field evolution in the early universe have, up to now, only been treated classically. This paper has shown that the extension of this classical treatment leads to fluctuation-dissipation dynamics.  In an earlier paper it was shown how fluctuation-dissipation effects would also extend the treatment of the Universe evolution in the general hot big bang regime \cite{Bastero-Gil:2014jsa}. Combined these papers provide an extended dynamical framework to examine key unsolved problems of the early universe.

The underlying principle behind both these papers is the same as in the original warm inflation work -  that many of the most fundamental quantities measured in cosmology, those associated with some underlying field dynamics, only provide a coarse-grained information about that dynamics and not necessarily about its microphysical properties. This is the concept that separates the warm and cold paradigms of inflation. In cold inflation, the observables such as the index and bispectrum, are interpreted to probe precise information about an underlying classical dynamics with quantum fluctuations superposed upon it. The uncertainty in this semiclassical approach is only that associated with the quantum mechanical uncertainty. In contrast warm inflation goes further to a full statistical state where these observables are interpreted to provide only coarse-grained information about the underlying fundamental dynamics. There can be many statistical dynamical realizations of the coarse-grained dynamics.  One common feature is the presence of energy fluxes amongst the coarse-grained cells, often related through the underlying dynamics. The fluctuation-dissipation relations examined in this paper are one, perhaps most common, example of such relations, and these can provide observable, testable, consequences.

Several problems in cosmology today need to be approached beyond the semiclassical approximation of thermal equilibrium dynamics, such as the generation of curvature and isocurvature perturbations, baryogenesis, leptogenesis, generation of dark matter, origin of cosmic magnetic fields, initial conditions of phase transitions, and dynamics during a phase transition or scalar field evolution. Adhering to just a thermal, semiclassical dynamical viewpoint of the underlying dynamics in the early universe can restrict the scope of theoretical investigation that is possible, and can lead to misleading directions of interpretation, such as doing elaborate model building where some simple statistical interpretation could actually bring the predictions in line with observation. This paper has highlighted these points and provided a methodology that can be used for such exploration along with several example applications.

%%%%%%%%%%%%%%%%%%%%%%%%%%%%%%%%%%%%%%%%%%%%%%%%%%%%%%%%%
%%%%%%%%%%%%%%%%%%%%%%%%%%%%%%%%%%%%%%%%%%%%%%%%%%%%%%%%%

\begin{acknowledgments}
We would like to thank Rudnei Ramos for useful discussions about this work. S.B.~and A.B.~are supported by the Science and Technology Facilities Council (United Kingdom). J.G.R.~is supported by the FCT grant SFRH/BPD/85969/2012 (Portugal), as well as partially supported by the grant PTDC/FIS/116625/2010 (Portugal) and the Marie Curie action NRHEP-295189-FP7-PEOPLE-2011-IRSES. J.G.R.~ and S.B would like to acknowledge the hospitality of the Higgs Centre for Theoretical Physics of the University of Edinburgh and the Departamento de F\'{\i}sica da Universidade de Aveiro during the completion of this work.  
\end{acknowledgments}

%%%%%%%%%%%%%%%%%%%%%%%%%%%%%%%%%%%%%%%%%%%%%%%%%%%%%%%%%%%%%%%%%%%%%%%%%%%%%%%%%%%%%%%%%%%%%%%%%%%%%%%%%%%%%%%%%%%%%%%%%%%%%%%%%%%%%%%%

\bibliographystyle{apsrev} 
                 
\bibliography{cpt.bib}

\end{document}